\pdfoutput=1
\documentclass[sigconf]{acmart}

\usepackage{listings}
\usepackage{placeins}
\usepackage{array}
\definecolor{planlistingbackground}{RGB}{248,249,251}
\definecolor{planlistingrule}{RGB}{211,216,223}
\definecolor{vocaspeech}{HTML}{1747E9}
\definecolor{vocacanvas}{HTML}{FF7475}

\lstdefinestyle{responseplan}{
  basicstyle=\ttfamily\footnotesize,
  backgroundcolor=\color{planlistingbackground},
  frame=single,
  rulecolor=\color{planlistingrule},
  framerule=0.35pt,
  framesep=5pt,
  aboveskip=0.55\baselineskip,
  belowskip=0.55\baselineskip,
  columns=fullflexible,
  keepspaces=true,
  showstringspaces=false,
  breaklines=true,
  breakatwhitespace=true,
  literate=
    {preface}{{{\color{vocaspeech}preface}}}{7}
    {speech}{{{\color{vocaspeech}speech}}}{6}
    {Speech}{{{\color{vocaspeech}Speech}}}{6}
    {canvas}{{{\color{vocacanvas}canvas}}}{6}
    {CanvasAction}{{{\color{vocacanvas}CanvasAction}}}{12}
    {representation.create}{{{\color{vocacanvas}representation.create}}}{21}
    {annotation.add}{{{\color{vocacanvas}annotation.add}}}{14}
    {attention.show}{{{\color{vocacanvas}attention.show}}}{14}
}

\AtBeginDocument{
  }

\setcopyright{none}
\renewcommand\footnotetextcopyrightpermission[1]{}

\begin{document}

\title{\textsc{VoCa}: Designing Speech--Canvas Interaction for
Voice-Based Conversational Agents}

\author{Yate Ge}
\affiliation{%
  \institution{Tongji University}
  \city{Shanghai}
  \country{China}
}
\email{geyate@tongji.edu.cn}

\author{Run Yuan}
\affiliation{%
  \institution{Tongji University}
  \city{Shanghai}
  \country{China}
}
\email{yuanrun0607@gmail.com}

\author{Yueran Qi}
\affiliation{%
  \institution{Grinnell College}
  \city{Grinnell}
  \state{Iowa}
  \country{United States}
}
\email{qisarah@grinnell.edu}

\author{Wenjie He}
\affiliation{%
  \institution{Tongji University}
  \city{Shanghai}
  \country{China}
}
\email{hewenjie\_0425@tongji.edu.cn}

\author{Jiaqi Mo}
\affiliation{%
  \institution{Tongji University}
  \city{Shanghai}
  \country{China}
}
\email{mojiaqi@tongji.edu.cn}

\author{Yangshuo Chen}
\affiliation{%
  \institution{Hunan University}
  \city{Changsha}
  \state{Hunan}
  \country{China}
}
\email{franchesca@hnu.edu.cn}

\author{Wenbin Zuo}
\affiliation{%
  \institution{Tongji University}
  \city{Shanghai}
  \country{China}
}
\email{wbzuo@tongji.edu.cn}

\author{Xiaohua Sun}
\authornote{Corresponding authors.}
\affiliation{%
  \institution{Southern University of Science and Technology}
  \city{Shenzhen}
  \state{Guangdong}
  \country{China}
}
\email{sunxh@sustech.edu.cn}

\author{Weiwei Guo}
\authornotemark[1]
\affiliation{%
  \institution{Tongji University}
  \city{Shanghai}
  \country{China}
}
\email{weiweiguo@tongji.edu.cn}

\author{Qi Wang}
\affiliation{%
  \institution{Tongji University}
  \city{Shanghai}
  \country{China}
}
\email{qiwangdesign@tongji.edu.cn}

\begin{abstract}
People write and sketch while speaking to explain, organize, and develop
content together. Inspired by these practices, we investigate how voice agents
can use a canvas alongside speech in multi-turn conversations with users. We
conducted a two-part formative study: an observational study of how pairs
coordinated speech and boardwork, followed by a design workshop that informed
a design space for speech--canvas interaction with voice agents. Building on
these insights, we developed \textsc{VoCa}, a voice agent that coordinates
speech with visual object creation, annotation, and attention guidance. A
five-day deployment with 18 participants examined usability, experiences of
speech--canvas interaction, patterns of use, and desired improvements.
Participants' experiences highlighted opportunities for speech--canvas
interaction in learning, work, and daily life, alongside challenges in coordinating
what agents say and show in ways users can follow and influence. These findings
inform how voice agents can use a canvas alongside speech in conversation.
\end{abstract}



\begin{teaserfigure}
  \centering
  \includegraphics[width=\textwidth]{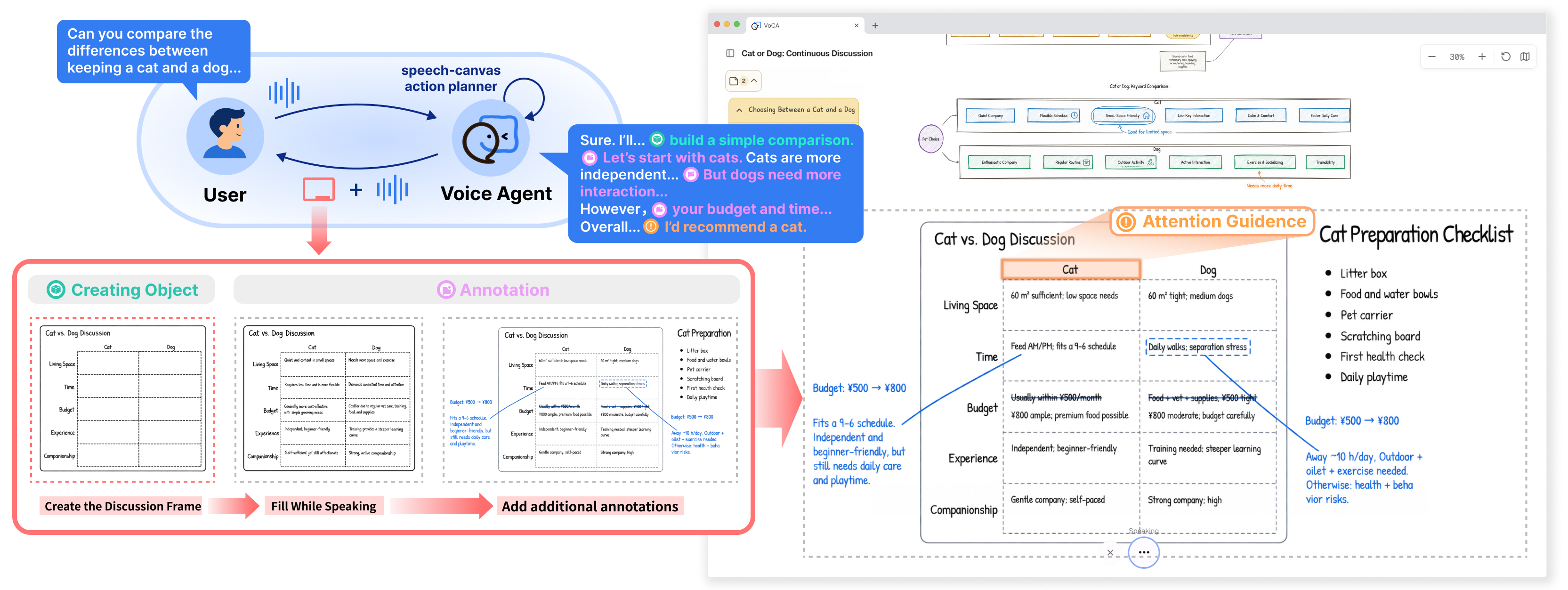}
  \caption{\textsc{VoCa} is a voice agent that coordinates speech with actions
  on the Canvas during multi-turn conversations. It creates and develops
  Visual Objects, adds Annotations, and directs attention to relevant content,
  allowing users and the agent to refer to, revisit, and
  continue developing shared content across turns.}
  \Description{A left-to-right overview of a cat-versus-dog discussion. At
  upper left, the user asks for a comparison and the voice agent jointly plans
  speech and Canvas actions. A three-step sequence below shows the agent
  creating an empty comparison frame, filling it while speaking, and adding
  annotations. At right, the resulting Canvas contains a cat-versus-dog
  comparison and a cat preparation checklist; focus cues connect the spoken
  explanation to selected cells and notes.}
  \label{fig:voca-teaser}
\end{teaserfigure}

\maketitle
\pagestyle{plain}

\section{Introduction}
\label{sec:introduction}

Voice-based conversational agents allow people to interact with computational
systems through spoken exchanges. With agents powered by large language
models, these exchanges can extend across turns as users ask follow-up
questions, seek clarification, and develop ideas \cite{mahmood2025voice}.
Voice products such as OpenAI's ChatGPT Voice\footnote{\url{https://help.openai.com/en/articles/20001274}}
and Google's Gemini Live\footnote{\url{https://gemini.google/overview/gemini-live/}}
support real-time spoken exchanges with natural-sounding speech and
user interruptions. Research on full-duplex dialogue and
interruption-aware voice agents likewise shows a shift beyond rigid,
turn-by-turn exchanges \cite{veluri2024synchronous,liu2025bargein}.
Recent systems have applied voice agents to learning, personal record keeping,
and multi-step interaction with graphical interfaces
\cite{openmaic2026,hoefer2025telltime,vu2024gptvoicetasker}. Together, these
developments create opportunities for sustained voice-based interaction across
learning, work, and daily life.
However, as such conversations accumulate information across turns, speech alone
provides few stable handles for inspecting relationships, referring back
precisely, or revising earlier content. Studies of speech-based thought
externalization report a related difficulty: extended spoken material and its
linear transcript can be difficult to review, reorganize, and
synthesize~\cite{li2026orality}.

In human conversation, people often combine speech with writing, sketching,
and pointing to explain ideas and develop understanding together. External
representations can make relationships explicit and support reasoning
\cite{scaife1996external}, while shared visual materials help people
establish what they are referring to \cite{gergle2013visual}. Boardwork
illustrates how spoken explanations and visual representations develop
together over time \cite{tan2023representations}. These practices suggest
an opportunity for voice agents as they engage in conversations that involve
connecting ideas, clarifying relationships, and working through questions
across turns. A canvas could make relevant content visible for users and
agents to refer to as they speak, and allow that content to develop through
subsequent explanations, questions, and responses. For example, a diagram
introduced during an explanation could become the focus of a follow-up
question, prompting the agent to point to or annotate the relevant part.

Prior work has explored visual feedback about a voice agent's state and
processing \cite{oh2024voicon,kirmayr2026intermediate}, grounding spoken
references in visible objects \cite{lee2024gazepointar,huynh2025spatial},
contextual visual aids and co-speech drawing
\cite{liu2023visualcaptions,ge2025gencomui,rosenberg2024drawtalking,
liu2026amiwrite}, and interactive visual interfaces generated for
language-model interaction
\cite{chen2026generative,jiang2023graphologue,cao2025malleable}.
These efforts establish several ways speech and visual interfaces can
complement one another. We investigate how a voice agent's decisions about
what to say, show, and change on a canvas shape users' opportunities to
understand content, express their needs, and develop ideas with the agent.
This involves both coordinating the agent's speech and canvas actions and
enabling users to influence subsequent turns through speech and visible
content. We examine these design choices and their use in learning, work, and daily life.

Supporting such interaction requires a voice agent to coordinate what it says
with what it shows or changes on the Canvas: \emph{whether} to use visual
support in a turn, \emph{what} to show or change, \emph{how} to represent
and organize the content, and \emph{when} those actions occur in relation to
speech. This coordination affects how an explanation unfolds, what users can
point to or ask about, and how the conversation proceeds from there.

To inform the design of this interaction, we conducted a formative study
comprising an observational study and a design workshop. In the observational study,
12 participants worked in six pairs, each completing three discussion tasks
using a digital whiteboard. One participant operated the digital whiteboard while both discussed the topic. We then conducted a design workshop in two sessions to
explore how voice agents could participate in conversations through speech
and canvas actions. From the workshop materials, we developed a design space
describing how canvas content is organized and developed, how users and agents
act on it, and how those actions relate to speech over time.

Informed by these findings and design explorations, we developed \textsc{VoCa}
(Figure~\ref{fig:voca-teaser}). Within this broader scope, \textsc{VoCa} focuses
on how a voice agent coordinates speech and canvas actions during interaction
with users. It creates visual objects, adds annotations, and directs attention
in coordination with speech, using the current Canvas to inform what it says
and does next. Users participate through speech,
with selections and the current
canvas view helping the agent interpret references to visible content.

To examine this interaction in use, we conducted an exploratory five-day
deployment study with 18 participants, who used \textsc{VoCa} for
self-selected everyday activities across devices. We collected interaction
logs, Session Reflections, a post-deployment questionnaire, and
semi-structured interviews. Combining descriptive analysis of the
questionnaire with qualitative analysis of participants' accounts and
interaction logs, we examined how participants assessed \textsc{VoCa}'s
usability and experienced its speech--canvas interaction, how they
used it in everyday activities, and what improvements they wanted.

This paper makes the following contributions:
\begin{itemize}
  \item A formative study revealing how people externalize content, link it to
  speech, and develop it as paired discussions unfold, together with a design
  space for speech--canvas interaction with voice agents;
  \item \textsc{VoCa}, a voice agent that uses speech and Canvas actions as
  coordinated ways of participating in conversation. It jointly plans spoken
  responses with visual-object creation, annotation, and attention guidance
  using the evolving Canvas as context, while users guide subsequent
  expression through speech and selection; and
  \item An exploratory five-day deployment study of \textsc{VoCa} with 18
  participants, evaluating its usability and examining how participants
  experienced its speech--canvas interaction and used it in everyday
  activities.
  Based on the findings, we discuss the value, application opportunities, and
  design implications of speech--canvas interaction with voice agents.
\end{itemize}

\section{Related Work}
\label{sec:related-work}

\subsection{Voice-Based Conversational Agents}

People use voice interfaces to access information and services within ongoing
activities, with interaction shaped by their practical needs and surrounding
social settings
\cite{ammari2019music,porcheron2018voice}.
With large language models, voice agents increasingly support open-ended,
multi-turn conversations, allowing users to elaborate on requests, ask
follow-up questions, and develop ideas through dialogue
\cite{mahmood2025voice,zhong2025textilebot}.
Advances in speech modeling also support interruption, backchanneling, and
overlapping speech, expanding how users and agents can participate in spoken
exchanges \cite{veluri2024synchronous,liu2025bargein}.
Systems such as OpenMAIC \cite{openmaic2026}, TellTime
\cite{hoefer2025telltime}, and GPTVoiceTasker \cite{vu2024gptvoicetasker}
illustrate applications in learning, personal record keeping, and multi-step
GUI tasks. Together, these developments broaden what voice agents can support
and the settings in which people use them. They also raise the question of how
visual interaction can become part of an ongoing voice conversation. Our study
addresses this question through speech--canvas interaction.

\subsection{Visual Interfaces in Voice Interaction}

Research comparing voice and screen-based interaction shows that modality
shapes how users carry out agent-assisted activities
\cite{reicherts2022good,chen2023screen}. Existing systems use visuals to
communicate a voice interface's conversational state
\cite{oh2024voicon}, expose an agent's intermediate processing
\cite{kirmayr2026intermediate}, or provide richer in-vehicle feedback
\cite{baghdadi2025hud}. Research also studies how proactive agents adapt
auditory and visual presentation to context \cite{lee2025sensible}.
Visible materials can also carry interaction content: shared materials can
maintain a record of the problem state
\cite{dillenbourg2006sharing}, while shared visual information can support
reference and grounding in collaborative tasks \cite{gergle2013visual}.
Recent systems ground spoken references in on-screen content
\cite{huynh2025spatial} or physical surroundings using gaze and pointing
\cite{lee2024gazepointar}. Systems also visually externalize their
interpretation of an ambiguous spoken query \cite{lee2026player}. Together,
this work shows how a visual surface can carry system feedback, task
information, and shared referents within voice interaction.

Speech also provides a way to act on graphical content. Systems coordinate
graphical interfaces and voice agents \cite{sorokin2025voicegui}, combine
speech and sketch to specify interface actions or create and revise graphical
worlds \cite{huang2025sketchgpt,rosenberg2024drawtalking}, or enable
speech-driven AR authoring with direct manipulation \cite{lee2025imaginatear}. Other work
enables voice control of mobile or scientific
interfaces \cite{vu2024gptvoicetasker,jia2025voice,hombeck2026vocalvis}, or
uses pointer context and low-interruption feedback to connect speech to ongoing
software work \cite{gmeiner2026pointaloud}. Architectural work connects
existing GUIs with conversational assistants \cite{vandam2025architecture}.
Studies of human boardwork examine how visual content takes shape during
spoken interaction. Speakers selectively write and connect content as an
explanation and others' contributions develop; speech, diagrams, spatial
organization, and pointing jointly shape the resulting representation
\cite{tan2021bansho,tan2023decision,tan2023representations}. These findings
align with research on multiple representations, which emphasizes
complementary functions and clear relations among representations
\cite{ainsworth2006deft,mayer2003nine}. Together, these studies direct attention
to both the visual form of content and how speech, spatial organization, and
pointing organize its use.

Recent work brings these concerns into co-speech visual generation. Visual
Captions proposed a design space for augmenting human--human verbal
communication, characterizing when, for whom, and in what form visuals are
presented, alongside communication setting, privacy, initiation, and
interaction modality \cite{liu2023visualcaptions}. GenComUI coordinates
speech and generated task visuals for iterative task confirmation and
refinement in robot programming \cite{ge2025gencomui}. AmIWrite studies
co-speech handwriting within mathematics tutoring \cite{liu2026amiwrite},
while OpenMAIC explores co-speech writing and drawing \cite{openmaic2026}.
Other systems use shared visual input for situated conversation
\cite{ho2025art}. Complementary work generates
whiteboard drawing sequences conditioned on narration
\cite{prasad2026whiteboard}. Emerging datasets and
systems also investigate temporally aligned sketch--speech input and proactive
diagram completion
\cite{shi2026drawing,ellawela2025whiteboards}.
Across these strands, prior work establishes multiple roles for visual
interfaces alongside speech. Our study focuses on how a voice agent uses speech
and a Canvas together in an ongoing conversation, and how users participate in that
conversation.

\subsection{Generative Visual Interfaces for Human--AI Interaction}

Large language models enable interfaces to be generated in response to users'
requests and context \cite{chen2026generative}. Generated responses range from
interface components \cite{livekit2026genui} to problem representations that
learners can inspect and annotate during tutoring, as in Interactive Sketchpad
\cite{lee2025sketchpad}. Other systems transform model output into
entity--relation diagrams, multilevel spatial structures, and organized design
alternatives \cite{jiang2023graphologue,suh2023sensecape,suh2024luminate}, or
make conversational goals and branching histories visible across turns
\cite{coscia2025ongoal,amin2026conversations}. These systems show how generated
interfaces can organize content, expose relationships, and provide entry
points for subsequent interaction.
Generative interfaces can also remain malleable after their initial creation.
Jelly uses an evolving task-driven data model to preserve data across interface
changes, DynaVis generates persistent controls for visualization editing, and
other systems maintain correspondences between requirements, GUI prototypes,
programs, and diagrams
\cite{cao2025malleable,vaithilingam2024dynavis,kretzer2025closing,ge2024cocobo}.
Data Player further demonstrates explicit coordination between narration and
visual animation in generated presentations \cite{shen2024dataplayer}.

Other work treats the canvas as a space in which people and agents
develop content over time. Orality extracts and organizes a user's spoken
thoughts on a malleable semantic canvas for individual reflection; CoMAP
supports learning design in a shared graph workspace; and the open-source Huabu
project brings materials and agents into a persistent workspace
\cite{li2026orality,li2026comap,microsoft2026huabu}. This line of work shows how
generated visual content can organize interaction and continue developing over
time. Our study examines the role of such content within ongoing interaction
between users and a voice agent.

\section{Formative Study}
\label{sec:formative-study}

\begin{figure*}[t]
  \centering
  \includegraphics[width=\textwidth]{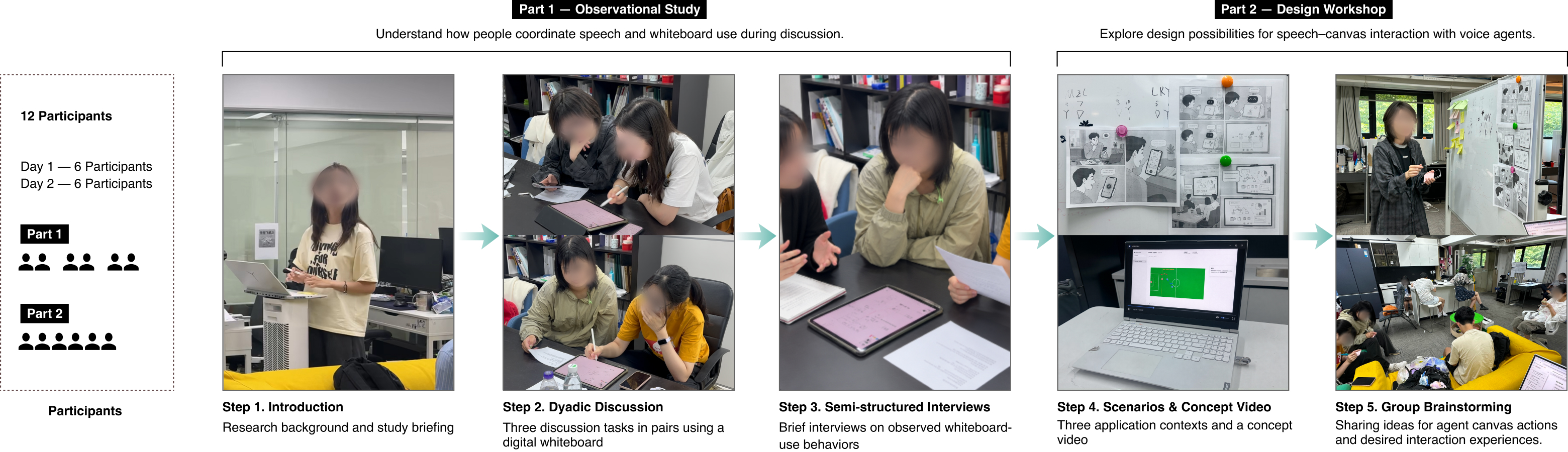}
  \caption{Overview of the formative study, conducted over two days with six
  participants per day. Part~1 comprised an introduction, three discussion
  tasks completed in pairs on a shared digital whiteboard, and brief
  semi-structured interviews about observed whiteboard use. Part~2 introduced
  three application contexts and a concept video, followed by a six-person
  group discussion of possible agent canvas actions and desired interaction
  experiences.}
  \Description{A left-to-right photographic flowchart shows five study steps:
  introduction, dyadic discussion, semi-structured interviews, scenarios and
  concept video, and group brainstorming. Brackets group the first three steps as
  an observational study and the last two as a design workshop.
  A side panel shows twelve participants across two days, with three pairs
  in the first part and one group of six in the second part on each day.}
  \label{fig:formative-study-procedure}
\end{figure*}

To inform the design of speech--canvas interaction with voice agents, we
examined how people combined speech and whiteboard use during paired
discussions. We focused on what participants externalized, how board actions
accompanied speech, and how existing content supported later discussion.
A subsequent design workshop explored possibilities for users and voice
agents to interact through speech and a canvas.
Figure~\ref{fig:formative-study-procedure} summarizes the procedure.

\subsection{Procedure}
\label{sec:formative-analysis}

We recruited 12 university students across two days (six per day). All
provided informed consent and received RMB~100 each. In Part~1,
each pair completed three discussion tasks: explaining a topic, developing
a solution or proposal, and making a decision or plan, yielding 18 sessions.
Task sheets specified goals and suggested topics; pairs could choose their
own topics within each task type. Both participants spoke, while one operated
the digital whiteboard. We recorded speech and board activity and conducted
brief follow-up interviews about participants' reasons for their actions.

In Part~2, each group of six joined a one-hour design workshop. We introduced
three application contexts: a meeting-room interactive whiteboard, a desktop
educational robot, and a mobile app with a voice agent. A concept
video illustrated an agent using a canvas while discussing soccer rules
with a user. Participants recorded ideas for agent canvas actions and desired
interaction experiences on sticky notes and explained them to the group.

Two researchers coded speech and board actions in the Part~1 videos,
grouping similar observations into descriptive codes and reconciling their
interpretations with reference to the interviews
(Appendix~\ref{app:formative-coding}). For Part~2,
we used affinity diagramming to group ideas by meaning, interpreting the
sticky notes alongside participants' explanations and using these groups
to inform our synthesis of the design space.

\subsection{Findings from the Observational Study}
\label{sec:formative-findings}

\subsubsection{\textbf{What Did Participants Put on the Board, and Why?}}
Participants selected what to draw alongside speech. In a follow-up interview,
one participant explained that they used speech or gestures for simple
concepts but began drawing when describing an entire media-streaming process
that they found difficult to convey through gestures alone. Other episodes
showed how selected information took visual form: in a snowboarding
explanation, the speaker sketched a board and its bindings, then added edge
labels and movement paths. In a scheduling discussion, the
board operator recorded dates and commitments while their partner described
them, adding stars to mark obligations. These marks gave the visible content
different roles: a structure to explain, details to locate within it, or
constraints to retain while considering a plan. The nine board-action codes
in Table~\ref{tab:formative-board-actions} describe these uses and the other
actions observed in the study.

\subsubsection{\textbf{How Did Speech and Board Actions Coordinate?}}
Writing or drawing accompanied related speech in all 18 sessions. Participants
also referred to existing content through speech and pointing in 17 sessions,
and questions or feedback prompted visual updates in 17
(Table~\ref{tab:formative-coordination}). In a guitar explanation, the speaker
indicated strings on the drawing and added an arrow while explaining their
differences. During a card-game explanation, a question about which roles
were publicly known led the speaker to add a visibility label above an
existing role name. Board actions thus accompanied both the speaker's account
and their responses to the partner, linking an explanation to a particular
part of the visible content. Coordination also included other local
sequences, such as drawing a framework before explaining it, which occurred
in six sessions.

\subsubsection{\textbf{How Did Participants Revisit and Develop Existing Content?}}
In the card-game session, a circular arrangement initially showed players'
turn order. Later, the partner reviewed the notes and asked about distance
rules; the speaker returned to the circle and annotated the distance between
two positions. The same structure now served a further explanation. Across
the study, participants revisited earlier content in 12 sessions.

More substantial reorganization accompanied a change in discussion stage in
ten sessions. In a scheduling session, participants first listed commitments
and marked their priorities. They subsequently drew a timeline placing work,
a wedding, and possible travel dates together, then produced a numbered plan
that separated the proposed itinerary from a gathering whose timing still
needed discussion. This sequence moved from collecting constraints to
considering their temporal relationships and consolidating an arrangement.
It illustrates how continued board use involved changing the organization
and purpose of visible content as well as adding information.

\subsection{A Design Space for Speech--Canvas Interaction with Voice Agents}
\label{sec:formative-design-space}

\begin{figure*}[t]
  \centering
  \includegraphics[width=\textwidth]{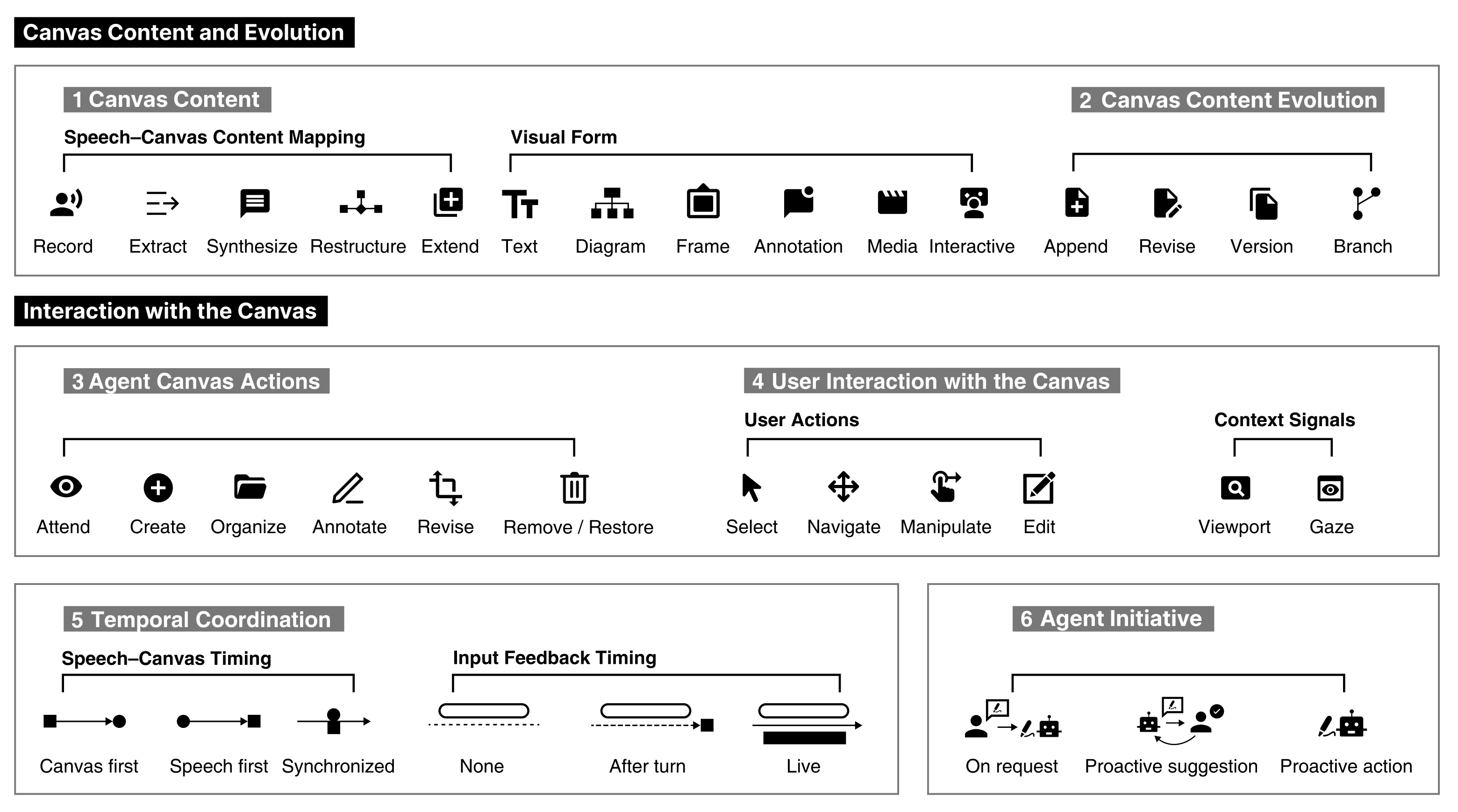}
  \caption{Design space for speech--canvas interaction with voice agents informed by the
  design workshop. D1--D2 describe canvas content and its evolution;
  D3--D6 describe agent actions, user interaction, temporal coordination,
  and agent initiative. Values within a dimension
  represent alternatives or capabilities that a system may combine.}
  \Description{A diagram groups six dimensions into Canvas Content and
  Evolution and Interaction with the Canvas. Canvas Content includes five
  speech--canvas content mappings: record, extract, synthesize, restructure,
  and extend; visual forms include text, diagram, frame, annotation, media,
  and interactive components. Canvas Content Evolution includes append,
  revise, version, and branch. Agent Canvas Actions include attend, create,
  organize, annotate, revise, and remove or restore. User Interaction with
  the Canvas distinguishes user actions---select, navigate, manipulate, and
  edit---from context signals---viewport and gaze. Temporal Coordination
  includes canvas-first, speech-first, and synchronized speech--canvas timing,
  and input feedback that is absent, appears after the user's turn, or updates
  live. Agent Initiative includes acting on request, proactively suggesting
  an action for user acceptance, and proactively acting without per-action
  confirmation.}
  \label{fig:speech-canvas-design-space}
\end{figure*}

Informed by the design workshop, we developed a design space for speech--canvas
interaction with voice agents (Figure~\ref{fig:speech-canvas-design-space}).
Its six dimensions describe choices about how content is expressed and developed,
how users and agents act on it, and how these actions unfold within the conversation.
The design space offers a basis for exploring and selecting interaction designs
suited to different conversational contexts.

\paragraph{\textbf{D1. Canvas Content.}}
This dimension concerns how information from a conversation becomes canvas
content and what visual form it takes. \emph{Speech--Canvas Content Mapping}
includes preserving the original expression (record), selecting key information
(extract), integrating information within or across turns (synthesize), making
relations or processes explicit (restructure), and contributing information
beyond what has been spoken (extend). These mappings can work together:
selected information can also be organized into a relational structure.
\emph{Visual Form} concerns how the result is expressed. Text carries verbal
content; diagrams express relationships; frames provide structures that can be
filled over time; and annotations attach emphasis, comments, or status to
existing content. Media provides images, audio, or video, while interactive
components respond to user operations through changes in state.

\paragraph{\textbf{D2. Canvas Content Evolution.}}
Workshop proposals included retaining intermediate plans for later comparison
and periodically consolidating content as ideas were revised or rejected.
This dimension describes how canvas content develops over time. Appending adds content
while retaining what is already present. Revising updates existing content or
its structure. Versioning retains earlier states for later review, whereas
branching allows alternatives to develop separately from a common starting
point. These choices affect whether users encounter an updated account, a
history of changes, or parallel directions they can continue exploring. They
can be combined: a branch can accumulate additions and retain versions of its
revisions.

\paragraph{\textbf{D3. Agent Canvas Actions.}}
This dimension describes the actions available to an agent. An agent can
direct attention by pointing to content or bringing it into view (attend),
create content, and connect, group, or arrange it (organize). It can add marks
or comments to existing content (annotate), modify the content itself (revise),
and remove content or restore an earlier state. These actions determine how an
agent can use both new and existing canvas content during a conversation.
D3 describes what actions are taken on canvas content, while D2 describes
how changes are retained and developed over time.

\paragraph{\textbf{D4. User Interaction with the Canvas.}}
Users can participate through \emph{User Actions} and \emph{Context Signals}.
Selection identifies an object, passage, or region; navigation changes the
view; manipulation moves or resizes objects or operates interactive components;
and editing adds, changes, or removes content. These actions let users refer
to material, browse it, and contribute directly. Context signals provide
additional information for the agent: the viewport indicates what is visible,
while gaze can indicate what the user is looking at. A system can use these
signals alongside speech to interpret references without requiring an explicit
selection each time.

\paragraph{\textbf{D5. Temporal Coordination.}}
Timing preferences varied across workshop notes: one requested a completed
canvas before explanation, while another proposed speaking and drawing
together. We distinguish two timing choices.
\emph{Speech--Canvas Timing} concerns the relation between an agent's utterance
and a related canvas action: the action can precede the speech (canvas first),
follow it (speech first), or occur at the corresponding moment in the utterance
(synchronized). Different parts of a conversation can use different arrangements.
\emph{Input Feedback Timing} concerns feedback on the user's input. A system
may provide none, present a record or interpretation after the user's turn,
or update feedback live while the user speaks. Live feedback can also be
refined after the turn ends.

\paragraph{\textbf{D6. Agent Initiative.}}
An agent can act on the canvas in response to an explicit user request
(on request), independently propose an action and wait for user acceptance
(proactive suggestion), or decide and act without a separate request or
confirmation for that action (proactive action). These arrangements distinguish
delegated execution, opportunities to review suggestions, and agent action
without prior review. Workshop proposals ranged from background organization
with automatic highlighting at key discussion points to presenting proposed
edits as annotations for user review. A system can combine different
arrangements across actions.

\subsection{Summary}

The observations provide design grounding for selective externalization,
coordination with speech, and continued use of existing content, consistent
with prior accounts of boardwork and visual aids in communication
\cite{tan2023representations,ge2025gencomui}. Workshop proposals informed our
synthesis of the design space for speech--canvas interaction with voice agents.
Together, these informed the design goals for
\textsc{VoCa} in the next section.

\section{System Design}
\label{sec:system}

\subsection{Design Goals}

Drawing on the formative findings and design space
(Section~\ref{sec:formative-study}), we formulated three design goals for
\textsc{VoCa}.

\paragraph{\textbf{DG1: Externalize Content in Forms Suited to the Conversation.}}
Participants selectively recorded information and used different visual structures
to support explanation, comparison, and planning. These practices motivate
selecting what to externalize and how to represent it according to the needs of
the conversation. The content mappings and visual forms in the design space offer
possibilities for this selection. We aim for the agent to use forms that help
users inspect relationships, compare ideas, or retain important contributions.

\paragraph{\textbf{DG2: Coordinate Speech and Canvas Actions within the Conversational Flow.}}
We aim for Speech and Canvas actions to make complementary contributions that
users can follow as a conversation unfolds. In the observational study, participants
coordinated boardwork with speech and used pointing to connect their utterances
with visible content. The temporal coordination dimension helps frame how these
channels can unfold together. Our goal is to coordinate what the agent says with
what it makes visible or refers to, at moments that preserve the flow of conversation.

\paragraph{\textbf{DG3: Develop Canvas Content as the Conversation Unfolds.}}
Inspired by how people write and sketch while talking, we aim for the agent to
build on Canvas content as the conversation unfolds. New contributions and
revisions should respond to the evolving conversation, while preserving a visible
record of how the discussion develops.

\subsection{System Overview}
\label{sec:system-overview}

\begin{figure*}[t]
  \centering
  \includegraphics[width=\textwidth]{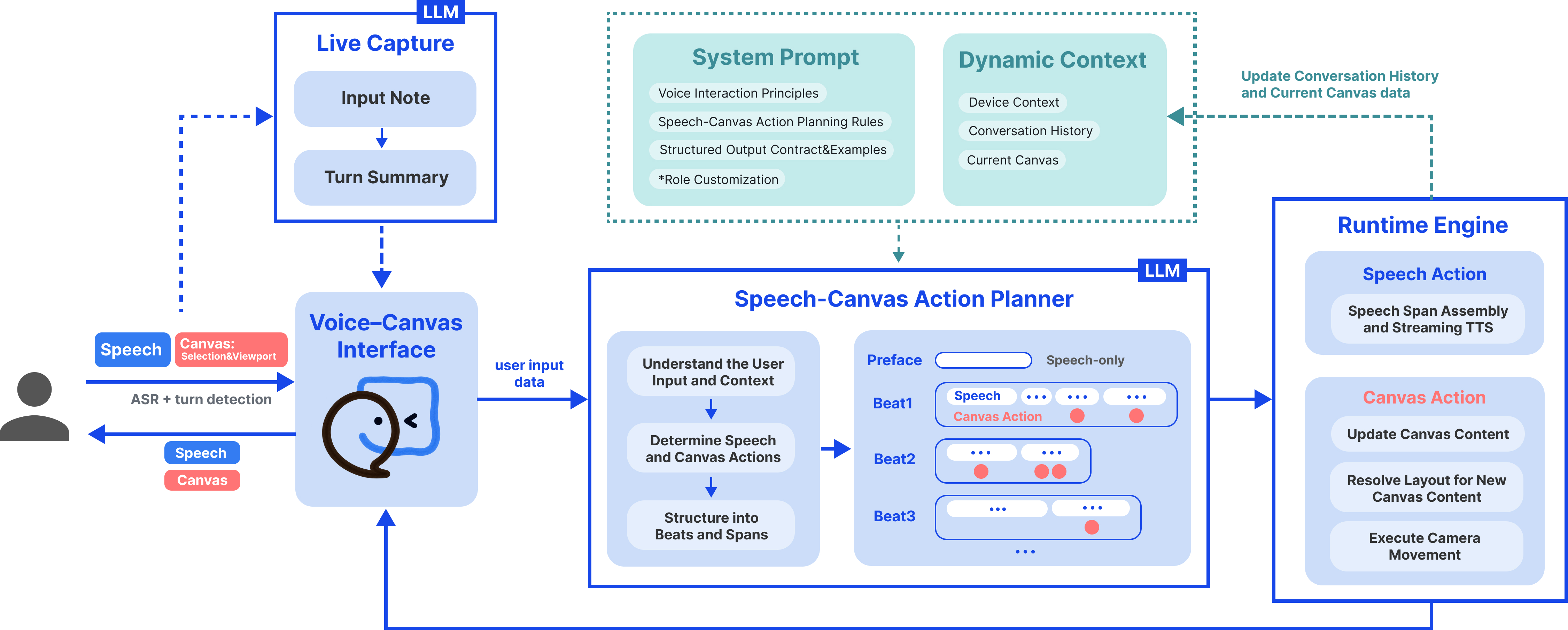}
  \caption{How \textsc{VoCa} works. The Speech--Canvas Action Planner uses the
  current input, System Prompt, and Dynamic Context to determine coordinated
  Speech and Canvas actions. A Speech-only Preface takes up the user's input;
  Beats then bind Canvas actions to Speech spans. The Runtime Engine assembles Speech spans for streaming text-to-speech and
  updates Canvas content, resolves layout for new Canvas content, renders focus
  cues, and executes Camera movement. The current user input and committed
  Runtime results update Conversation History and the Current Canvas for
  subsequent turns.}
  \Description{From left to right, a user speaks and may select or navigate the
  Canvas through the Voice--Canvas Interface. During speech, Live Capture
  incrementally updates a temporary Input Note and, when the turn ends,
  consolidates it into a user-attributed Turn Summary. A System Prompt and
  Dynamic Context, comprising Device Context, Conversation History, and the
  Current Canvas, inform the Speech--Canvas Action Planner. The Planner
  interprets the input, determines Speech and Canvas actions, and structures
  them into a Speech-only Preface followed by Beats containing Speech spans
  with optional Canvas actions. The Runtime Engine assembles Speech spans for
  streaming text-to-speech and executes Canvas content updates, layout, focus
  cues, and Camera movement. The current user input and committed Runtime
  results update Conversation History and the Current Canvas, which combine
  with Device Context to form the Dynamic Context for subsequent turns.}
  \label{fig:voca-system-architecture}
\end{figure*}

To realize these goals, we developed \textsc{VoCa}, a voice agent that
coordinates speech and Canvas actions in ongoing conversations with users.
Figure~\ref{fig:voca-system-architecture} presents the system architecture of
\textsc{VoCa}. Live
Capture provides visual feedback while the user speaks. The Speech--Canvas
Action Planner produces a spoken Preface and organizes the subsequent
Speech and Canvas actions into Beats and spans. The Runtime Engine then presents these actions and
commits their results to support subsequent turns.

\subsubsection{\textbf{The \textsc{VoCa} Interface}}
\label{sec:voca-interface}
\textsc{VoCa}'s interface, illustrated in the teaser
(Figure~\ref{fig:voca-teaser}), centers on a \emph{Canvas}. A voice button
starts voice interaction and indicates the agent's current status. Users can
choose between real-time voice, which automatically detects turn endings and
spoken interruptions, and manual voice, in which they control recording and
submission through the button. Both modes use the same Speech--Canvas planning
and presentation mechanisms.
To foreground our investigation of proactive agent actions on the Canvas, we
make the agent the primary Canvas actor
(Section~\ref{sec:agent-canvas-actions}).
Users guide the interaction through speech and interact with the Canvas
itself only by panning, zooming, selecting elements, and moving whole Visual Objects. Alongside Speech,
their current selection and the Canvas content visible in the current viewport
provide interaction context, allowing the agent to interpret references in
relation to what the user has selected or is viewing. Users cannot directly
create or edit Canvas content or make fine-grained layout adjustments. A collapsible sidebar lets
users start a new Session or reopen previous Sessions from a history list.

While users speak, \emph{Live Capture} provides visible input feedback on the
Canvas. It incrementally updates an \emph{Input Note} with what they have expressed.
After the turn is finalized, the Input Note is consolidated into a
user-attributed \emph{Turn Summary} for later reference, without copying the
transcript (Figure~\ref{fig:voca-system-architecture}).

\subsubsection{\textbf{Speech--Canvas Action Planner}}
\label{sec:speech-canvas-planning}

The \emph{Speech--Canvas Action Planner} receives the current user input
alongside a static \emph{System Prompt} and \emph{Dynamic Context}
(Figure~\ref{fig:voca-system-architecture}). The \emph{System Prompt} combines
\emph{Voice Interaction Principles}, optional \emph{Role Customization},
\emph{Speech--Canvas Action Planning Rules}, and a \emph{Structured Output Contract
and Examples}. \emph{Dynamic Context} supplies the \emph{Conversation History},
\emph{Current Canvas}, and \emph{Device Context} for the current turn.

The Planner interprets the current input together with the conversation history
and existing Canvas to determine the contribution needed in this turn.
User requests to change Canvas content are handled through this planning
process, with the agent determining the appropriate actions under its
interaction principles and available Canvas capabilities.
Guided by the \emph{Speech--Canvas Action Planning Rules}, it coordinates the
spoken response with content selected for visual support. Speech provides a
natural conversational response, while the Canvas externalizes structures,
relationships, states, or concise records that benefit from inspection,
comparison, or further development. Speech can introduce, explain, or refer to
this content without reading it out in full. The agent can develop the current
Canvas through annotations and new visual objects connected to existing content,
allowing its visual contributions to build on the ongoing conversation (DG3).

The Planner expresses these coordinated decisions as a spoken Preface followed
by Beats containing Speech spans and associated Canvas actions. The
\emph{Preface} is a brief spoken response that takes up the user's input and,
when a Visual Object will be created, signals the upcoming action. It contains
no Canvas actions and can begin playing while the rest of the response is still
being generated. The subsequent response is organized into Beats, each forming
a self-contained spoken unit. Within each Beat, Speech spans may carry one or
more Canvas actions triggered at the onset of the corresponding spoken segment (DG2).
In the structure below, \texttt{preface} encodes the Preface, and
\texttt{beats} contains the ordered Beats (see
Appendix~\ref{app:planner-output-example} for a concrete example):

\begin{lstlisting}[style=responseplan,float=!htbp]
{
  preface: Speech_0,
  beats: [
    { spans: [
      { canvas: [CanvasAction_1], speech: Speech_1 },
      { speech: Speech_2 },
      { canvas: [CanvasAction_2, CanvasAction_3], speech: Speech_3 }
    ] },
    { spans: [
      { speech: Speech_4 },
      { canvas: [CanvasAction_4], speech: Speech_5 }
    ] },
    { spans: [
      { canvas: [CanvasAction_5, ...], speech: Speech_6 }
    ] },
    ...
  ] }
\end{lstlisting}

The \emph{Runtime Engine} receives this output, assembles Speech spans for streaming
text-to-speech, updates Canvas content, resolves layout for new content, renders
focus cues, executes Camera movement, and commits completed results.
Section~\ref{sec:system-implementation} describes this execution path in more
detail.

\subsection{Canvas Actions}
\label{sec:agent-canvas-actions}

\begin{figure*}[t]
  \centering
  \includegraphics[width=\textwidth]{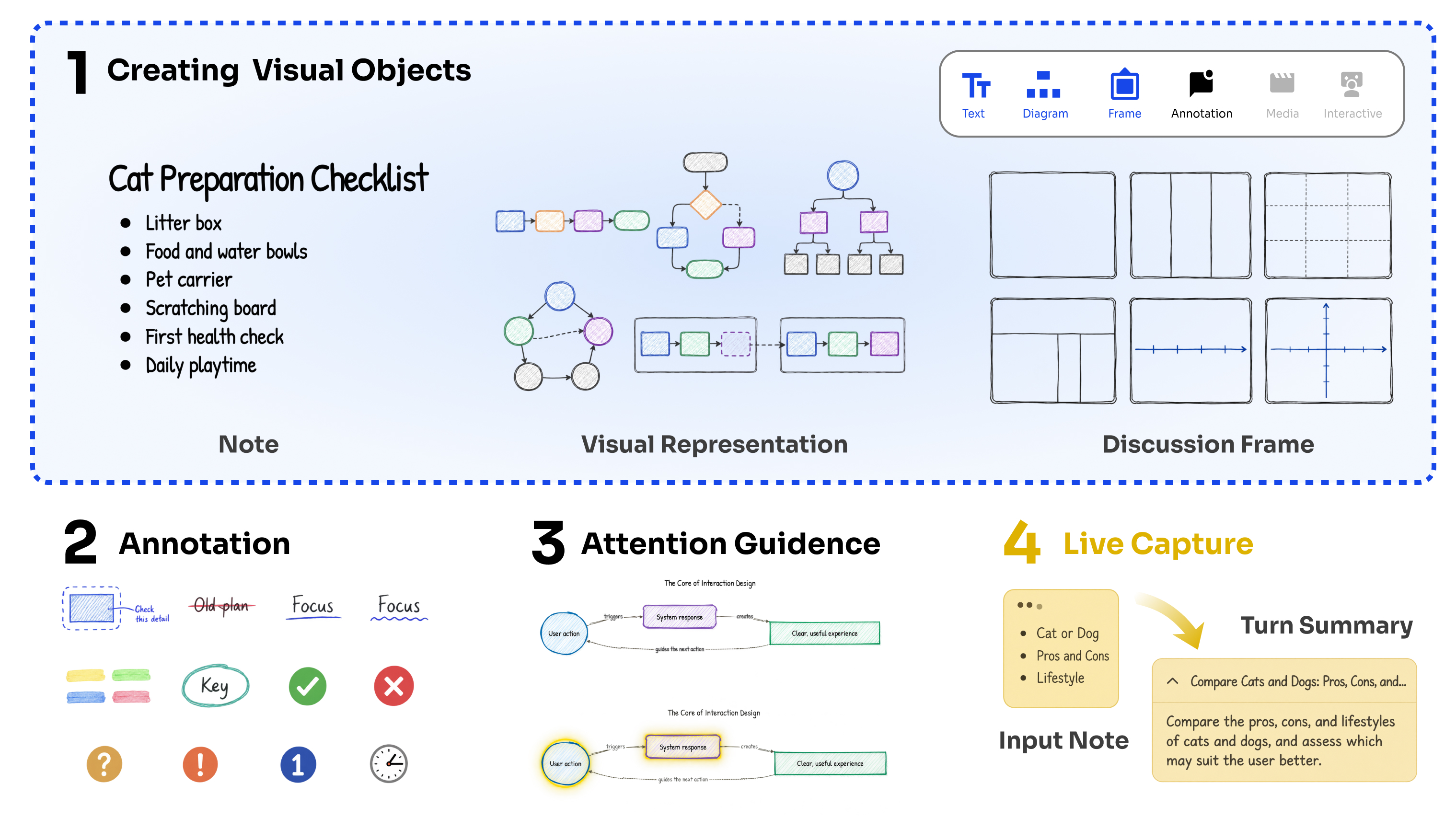}
  \caption{Canvas actions and Live Capture in \textsc{VoCa}. (1) Creating
  Visual Objects includes Notes, Visual Representations, and Discussion Frames.
  (2) Annotation adds persistent Comments, Badges, Visual Marks, and Frame Entries
  to existing content. (3) Attention Guidance uses temporary focus cues to point to current
  referents. (4) Live Capture incrementally updates an Input Note
  during speech and consolidates it into a user-attributed Turn Summary after
  the turn.}
  \Description{Four numbered panels summarize \textsc{VoCa}'s Canvas
  interaction. Panel 1 shows a Note, several Visual Representation structures,
  and six example Discussion Frame arrangements. An inset presents six visual-form
  options from the formative design space, identifying text, diagram, frame, and
  annotation as selected for \textsc{VoCa}, while media and interactive objects remain
  unselected. Panel 2 shows examples of Comments, Highlights, underlines, circles,
  Badges, and other persistent Annotation marks. Panel 3 contrasts an ordinary
  diagram with the same diagram brought into visual focus. Panel 4 shows an Input
  Note becoming a Turn Summary.}
  \label{fig:canvas-actions-live-capture}
\end{figure*}

Through the Canvas, \textsc{VoCa} creates Visual Objects, annotates existing
content, and guides users' attention as it creates, annotates, or refers to
Canvas content. It coordinates each action with what it says. To retain a visible
record as the conversation develops, the agent expresses revisions through
Annotations or new content, preserving committed contributions rather than
overwriting or deleting them (DG3). Figure~\ref{fig:canvas-actions-live-capture}
summarizes these three forms of Canvas action together with Live Capture.

\subsubsection{Creating Visual Objects}
\label{sec:creating-visual-objects}

To complement transient speech, \textsc{VoCa} selects among three types of
\emph{Visual Object} according to what should remain inspectable or open to
further construction (DG1; Figure~\ref{fig:canvas-actions-live-capture}, panel~1):

\begin{itemize}
  \item \textbf{Note.} A Note preserves a conversational contribution as a bounded
  piece of text when its wording or textual organization matters---for example, a
  definition, draft passage, outline, decision, or working list. It keeps this
  content visible as a shared textual artifact that
  participants can revisit or build on across turns. Specific text ranges remain
  addressable, allowing the agent to attach Annotations or direct Attention to
  particular wording while Speech introduces or discusses it.
  \item \textbf{Visual Representation.} When understanding depends on inspecting
  several parts and their relationships at once, a Visual Representation spatially
  externalizes a bounded understanding through visual structure,
  such as grouping, hierarchy, sequence, comparison, or explicit relationships, and
  leaves stable internal targets for later reference.
  \item \textbf{Discussion Frame.} When an organization has been established but its
  contents remain open to contribution, comparison, selection, or revision, a
  Discussion Frame provides stable, named places where contributions can be added,
  compared, and revisited across turns. Region partitions, Grids, Ordered axes, and
  Orthogonal planes organize these places, while target-dependent judgments or
  status are added as Annotations.
\end{itemize}

\paragraph{\textbf{Incremental layout.}}
New Visual Objects are placed incrementally according to their source relationships
with existing Canvas content. Each object's primary source guides its local
placement. References to internal elements favor nearby positions outside the
source object, while whole-object references follow the display's preferred growth
direction. Internal-element references produce visible connections; whole-object
relationships are retained without a line. Objects without a source continue near
the most recently added object. New placements use available space without
rearranging existing objects, making visible how contributions build on earlier
content (DG3).

\begin{figure}[t]
  \centering
  \includegraphics[width=\columnwidth]{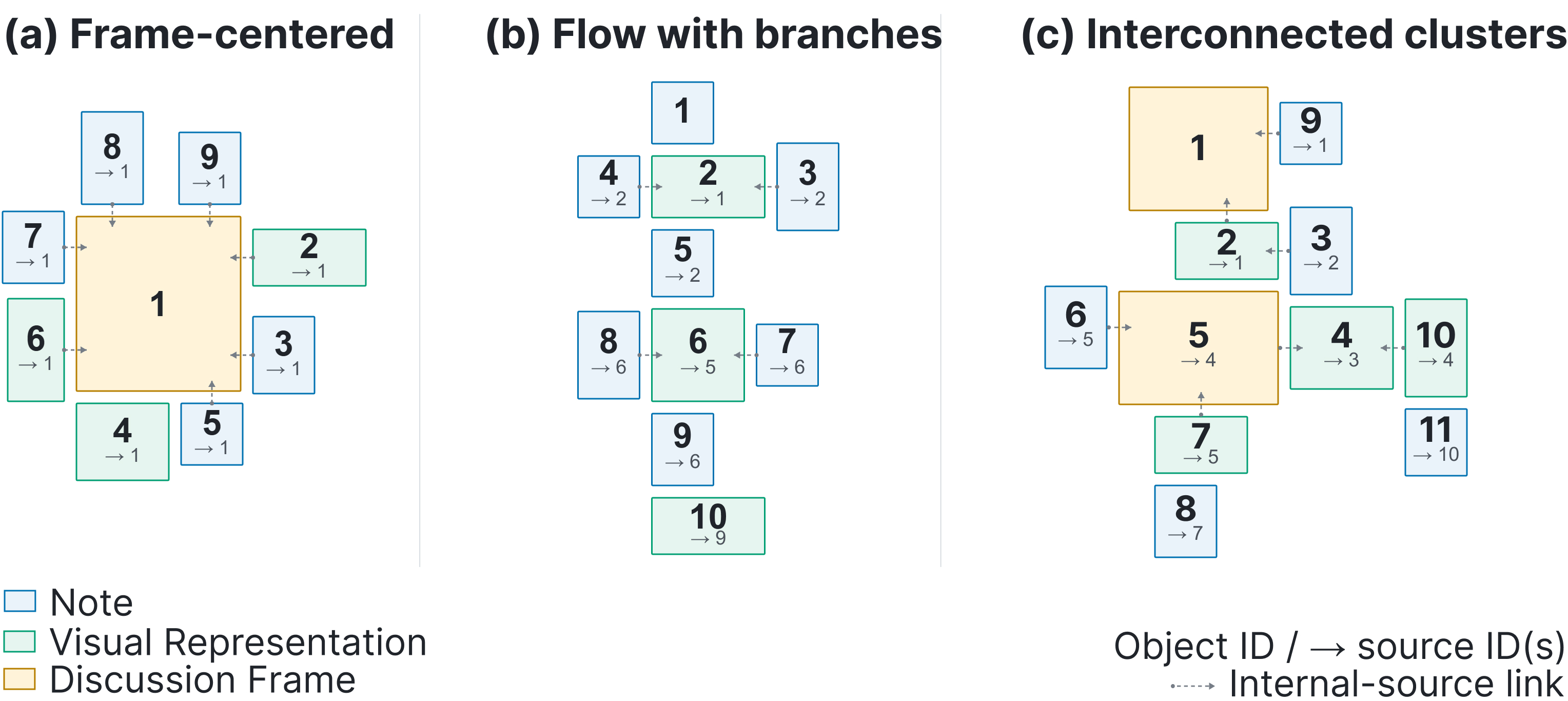}
  \caption{Illustrative spatial arrangements that can arise through incremental
  placement: (a) Frame-centered expansion, (b) flow with branches, and
  (c) interconnected clusters. Positions and sizes are schematic.}
  \Description{Three side-by-side layouts use blue for Notes, green for Visual
  Representations, and amber for Discussion Frames. Panel a has Frame 1 in the
  center surrounded by eight objects, all sourced from Frame 1. Objects 2 and 3
  are on the right, 4 and 5 below, 6 and 7 on the left, and 8 and 9 above.
  Seven objects have internal-source links; Representation 4 references the
  whole Frame and has no line.
  Panel b contains no Discussion Frames. After objects 1 and 2, Notes 3 and 4
  branch right and left from Representation 2, respectively. Note 5 continues
  below 2, followed by Representation 6, its right branch 7 and left branch 8,
  Note 9 below, and Representation 10.
  Panel c starts with Frame 1 and Representation 2 below it. Note 3 branches
  right from 2, and Representation 4 continues below 3. Frame 5 expands left
  from 4, followed by Note 6 to its left, Representation 7 below, and Note 8
  below 7. Note 9 later branches right from Frame 1. Representation 10 branches
  right from the earlier Representation 4, followed by Note 11 below 10.}
  \label{fig:incremental-layout}
\end{figure}

Figure~\ref{fig:incremental-layout} schematically illustrates possible spatial
arrangements that can arise as related objects are added over time: contributions
around a Discussion Frame, a sequence with parallel branches, and interconnected
clusters extending from earlier content.

\subsubsection{Annotating Existing Canvas Content}
\label{sec:annotation}

When new material derives its meaning from existing Canvas content, \textsc{VoCa}
adds it as an \emph{Annotation} rather than a new Visual Object. An Annotation either
attaches to a target as a Comment, Badge, or Visual Mark, or enters an established
Discussion Frame location as a Frame Entry. By tying new explanations, judgments,
or contributions to existing content, these additions let the agent develop what
has already been established across turns (DG3). They remain available for later
reference (Table~\ref{tab:annotation-forms}).

\begin{table}[H]
  \caption{Persistent Annotation forms and their roles in \textsc{VoCa}.}
  \label{tab:annotation-forms}
  \small
  \begin{tabular}{@{}>{\raggedright\arraybackslash}p{0.22\columnwidth}
    >{\raggedright\arraybackslash}p{0.34\columnwidth}
    >{\raggedright\arraybackslash}p{\dimexpr0.44\columnwidth-4\tabcolsep\relax}@{}}
    \toprule
    \textbf{Form} & \textbf{When used} & \textbf{Canvas result} \\
    \midrule
    \textbf{Comment} & Explanation, rationale, or question depends on a target.
      & Target-linked text with a lightweight leader. \\
    \addlinespace[3pt]
    \textbf{Badge} & Status or expressive cue should be recognizable at a glance.
      & Icon, number, or emoji on target. \\
    \addlinespace[3pt]
    \textbf{Visual Mark} & Existing content needs emphasis or a visible revision
      state. & Highlight, underline, circle, or strike on target. \\
    \addlinespace[3pt]
    \textbf{Frame Entry} & Contribution belongs in an established Frame location.
      & Compact text in a region, cell, axis, or plane position. \\
    \bottomrule
  \end{tabular}
\end{table}

Comments, Badges, and Visual Marks can apply to an entire Visual Object, an internal
element, a Frame region or Entry, or a specific text range within a Note; Frame
Entries instead occupy established Frame locations. Material that should stand on
its own or establish a new organization becomes a new Visual Object. Panel~2 of
Figure~\ref{fig:canvas-actions-live-capture} illustrates these forms on existing
Canvas content.

\subsubsection{Attention Guidance}
\label{sec:attention-guidance}

\emph{Attention Guidance} helps users connect spoken references with the
corresponding visible content (DG2).
When a target involved in the current Speech falls outside the readable viewport,
the Camera pans or zooms to keep its owning Visual Object fully visible. For an
explicit reference, a temporary focus cue marks the referenced Visual Object,
internal element, Frame location or Entry, or specific text range within a Note.
The cue remains only for the corresponding Speech and then fades, leaving committed
Canvas content and geometry unchanged (Figure~
\ref{fig:canvas-actions-live-capture}, panel~3).

User selection communicates ``I mean this'' and becomes context for the agent's
subsequent response; the agent's focus cue communicates ``this is what I am referring
to.'' The Camera remains still when the target is already readable. User-initiated
panning or zooming temporarily pauses automatic Camera following, allowing users to
inspect the Canvas without competing movement.

\subsection{Usage Scenario}
\label{sec:usage-scenario}
\begin{figure*}[t]
  \centering
  \includegraphics[width=\textwidth]{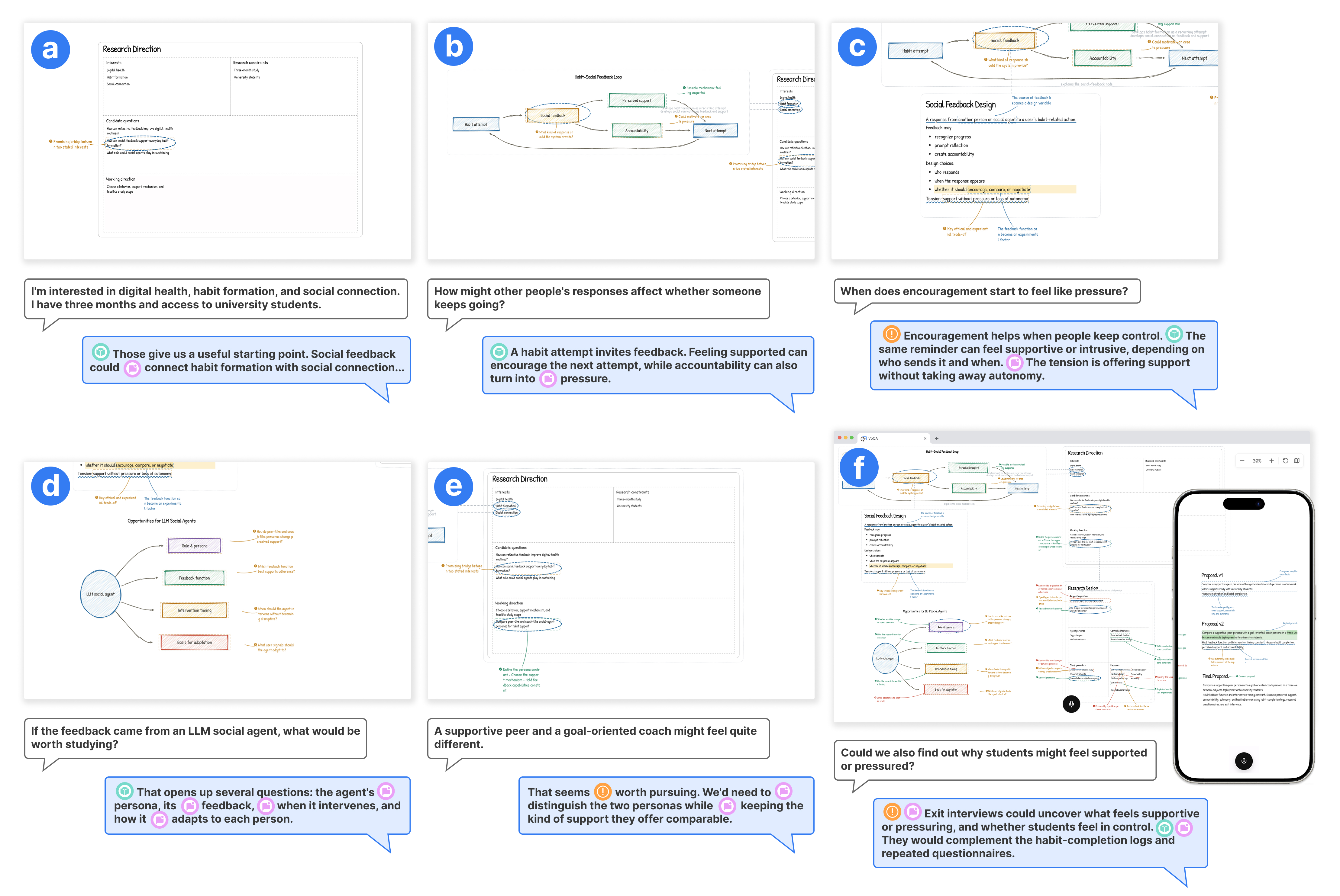}
  \caption{Six moments from a multi-turn discussion with \textsc{VoCa}. Xiaolin
  and the agent move from framing three interests and project constraints, through
  tracing a social-feedback loop and comparing research opportunities, to selecting
  a persona-based direction and turning it into a proposal. Panels (a)--(f) proceed
  from top left to bottom right; in panel (f), Xiaolin continues the same Canvas on
  her phone.}
  \Description{Six panels in two rows of three combine successive Canvas states
  with user and agent speech. Panels a through c show a Research Direction
  Frame, a habit--social-feedback loop, and a Social Feedback Design Note.
  Panels d through f show an opportunity representation for large-language-model
  social agents, the selected direction added to the Research Direction Frame,
  and an expanded Canvas containing the developed research proposal. The final
  panel also shows the proposal Notes continuing in a portrait phone view.}
  \label{fig:multiturn-usage-scenario}
\end{figure*}

Figure~\ref{fig:multiturn-usage-scenario} follows Xiaolin, a graduate student,
as she develops a thesis proposal with \textsc{VoCa} across six moments. As the
research direction narrows from broad interests to a concrete study, the agent
creates, extends, annotates, and revisits Canvas content. The same Canvas remains
available when Xiaolin continues on her phone.

\noindent\textbf{(a) Framing interests and constraints.}
Xiaolin wants to develop a research direction around digital health, habit
formation, and social connection within three months, preferably with university
students. As she speaks, Live Capture reflects these key terms in an Input Note.
The agent then creates a \emph{Research Direction} Frame, adds the three interests
and project constraints as Frame Entries, and uses them to suggest promising
connections and a feasible scope (Figure~\ref{fig:multiturn-usage-scenario},
panel~(a)).

\noindent\textbf{(b) Tracing a feedback loop.}
Xiaolin chooses to explore how habit formation and social connection might relate.
The agent marks the corresponding Frame Entries and constructs a feedback loop from
a habit attempt, through social feedback and perceived support or accountability,
to the next attempt. Additional Annotations preserve both the promise of
encouragement and the risk of pressure (Figure~
\ref{fig:multiturn-usage-scenario}, panel~(b)).

\noindent\textbf{(c) Unpacking social feedback design.}
When Xiaolin asks what ``social feedback design'' means, the agent uses a focus cue
to point to the social-feedback node and creates a nearby Note. The Note explains
that feedback can recognize progress, prompt reflection, and create accountability;
a Comment preserves the central tension of offering support without pressure or
loss of autonomy (Figure~\ref{fig:multiturn-usage-scenario}, panel~(c)).

\noindent\textbf{(d) Expanding design opportunities.}
Xiaolin next asks what concrete questions she could study through large language
model social agents. Drawing on the feedback loop and Note, the agent creates a
Visual Representation of four opportunity areas: role and persona, feedback
function, intervention timing, and basis for adaptation. Brief Annotations make
the question within each area available for further discussion (Figure~
\ref{fig:multiturn-usage-scenario}, panel~(d)).

\noindent\textbf{(e) Choosing a persona direction.}
Xiaolin decides to study how different social-agent personas support habit formation.
The agent marks the selected opportunity, adds it to the \emph{Research Direction}
Frame as a Frame Entry, and attaches a Comment preserving three open questions: how
to distinguish the personas, which support mechanism to examine, and which feedback
features to hold constant (Figure~\ref{fig:multiturn-usage-scenario}, panel~(e)).

\noindent\textbf{(f) Finalizing a study across devices.}
After further turns set out a \emph{Research Design} Frame and two
proposal drafts, Xiaolin continues the same Canvas on her phone and asks \textsc{VoCa} to
finalize the proposal while adding a qualitative component. The agent adds exit
interviews to the \emph{Research Design} Frame and creates a \emph{Final Proposal}
Note: a three-week between-subjects deployment comparing supportive-peer and
goal-oriented-coach personas while holding feedback function and intervention timing
constant. The earlier \emph{Proposal v1} and \emph{Proposal v2} Notes remain
available for comparison (Figure~\ref{fig:multiturn-usage-scenario}, panel~(f)).

\subsection{System Implementation}
\label{sec:system-implementation}

We implemented \textsc{VoCa} as a web application. The frontend and backend share
Zod schemas for Planner output and Canvas state, keeping model-produced actions
and Runtime operations within the same contract.
The backend persists the Conversation, Visual Objects, and geometry
for each Session, so desktop and mobile devices load the same Canvas and adapt
only viewport and Camera. A Web App Manifest also allows installation as a
progressive web app. Model calls, speech recognition, and speech synthesis use
the models in Table~\ref{tab:implementation-models}.

\begin{table}[t]
  \caption{Models used in the \textsc{VoCa} prototype.}
  \label{tab:implementation-models}
  \small
  \begin{tabular}{@{}p{0.30\columnwidth}p{0.64\columnwidth}@{}}
    \toprule
    \textbf{Function} & \textbf{Model} \\
    \midrule
    Real-time voice (turn detection, transcript, interruption) & Qwen-Audio 3.0 Realtime Flash \\
    Streaming transcription (Live Capture, manual voice) & Qwen3-ASR-Flash-Realtime \\
    Live Capture & DeepSeek V4 Flash (thinking disabled; temperature 0) \\
    Planner & GPT-5.6 (reasoning disabled; temperature 0) \\
    Speech synthesis & Qwen-Audio 3.0 TTS Flash (word-level timestamps) \\
    \bottomrule
  \end{tabular}
\end{table}

\textbf{Dynamic Context.} Conversation History contains the utterances
that were actually spoken and short records of Canvas changes that executed
successfully, so that later references can be resolved. Current Canvas is given
to the Planner as text rather than a screenshot: for each Visual Object, its
identifier, content or summary, source relations, Annotations, referable parts
(such as a Node or a paragraph), and position and size. The user's selection and
the device's Canvas dimensions accompany each turn.

\textbf{Runtime Execution.} The Runtime Engine handles the Planner's
streamed output one Beat at a time. Each complete Beat is validated for structure,
references, and dependencies before anything in it is played or executed. Its
Speech spans are joined in order into one utterance and synthesized by Qwen TTS. Canvas actions remain attached to the character positions of the Speech spans they
support; word-level timestamps from TTS map those positions to playback cues.
When a cue fires, the Runtime waits until the action's targets are
rendered, then executes it. On interruption, the Runtime stops playback and
cancels remaining execution, retaining what has already been spoken and
executed. If the model returns blank output or is unavailable before producing
any content, the Runtime retries once; otherwise the turn ends and the user can
resubmit. Deployment latency and failure rates are reported in
Section~\ref{sec:findings-rq1}.

\textbf{Canvas Rendering.} Each Visual Object has a specialized renderer. Notes use Markdown; images and raw HTML are
not rendered. Visual Representations use restricted Mermaid flowcharts:
Mermaid computes the layout, and the Runtime sanitizes
the SVG and exposes stable internal targets, so that later actions can
refer to individual parts of a diagram. Discussion Frames use a typed
structure that the Runtime compiles into a fixed scaffold, placing each
Entry at the Frame location the Planner specifies.

\section{User Study}
\label{sec:method}

We conducted an exploratory five-day deployment study to examine
\textsc{VoCa} in everyday use. Participants could use \textsc{VoCa} for their
own purposes across multiple devices, allowing us to learn from their experiences
and their expectations for using it in daily life. The study addressed three
research questions:
\begin{enumerate}
  \item[\textbf{RQ1.}] How do participants assess \textsc{VoCa}'s overall
  usability, and how do they experience its speech--Canvas interaction?
  \item[\textbf{RQ2.}] What do participants use \textsc{VoCa} for,
  and in what ways do they engage in speech--Canvas interaction with it?
  \item[\textbf{RQ3.}] Based on their deployment experience, what
  improvements do participants want from
  \textsc{VoCa}?
\end{enumerate}

\subsection{\texorpdfstring{Participants and Procedure}{Participants and Procedure}}

We recruited 19 participants through social media and snowball sampling.
All participants used LLM-based tools in their everyday activities and had
experience with voice interaction. Before enrolling, they were informed that the
study required using \textsc{VoCa} on each of five consecutive days.
One participant (P13) was excluded from the analysis for not meeting this
requirement, leaving 18 participants in the final analysis (10 women
and 8 men, aged 21--28 years; M = 22.61, SD = 1.97).
Participants were university students or
recent graduates from various disciplines, including design, computer science,
physics, and mathematics.
They were native Chinese speakers and used \textsc{VoCa} in Chinese.
All participants provided informed consent. Compensation was RMB 100 per participant.
The study was approved by our institution's ethics review process.

Before deployment, participants completed a questionnaire on demographics and
prior AI and voice-interaction experience, followed by a brief orientation with
guided practice using \textsc{VoCa}. The orientation included a demonstration
video and an introduction to \textsc{VoCa}'s basic functions and usage.
We deployed an online version of
\textsc{VoCa} and gave each participant an individual account, which they could
use to log in on different devices. We asked participants to use \textsc{VoCa} on
multiple devices where possible, and showed them how to add it to a phone's home
screen as a progressive web app. Real-time voice was enabled by default;
participants could switch to manual voice according to their needs.

During the deployment, participants used \textsc{VoCa} freely each day and were asked to converse over as many turns as possible and to end each
conversation by selecting \emph{Finish discussion} and submitting a brief Session
Reflection. The system did not enforce this step. Participants could contact us
with any questions during the deployment. At the end
of the deployment, they completed a questionnaire and an individual
semi-structured interview, which was audio-recorded.

\subsection{\texorpdfstring{Data Collection and Analysis}{Data Collection and Analysis}}

We collected a post-deployment questionnaire, interviews, Session
Reflections, and interaction logs. Study data were stored on a server
controlled by the research team.

\textbf{Questionnaire.} The post-deployment questionnaire comprised the System Usability Scale
(SUS)~\cite{brooke1996sus}, which gave a standardized overall assessment of
usability, and two sets of six study-specific items. The voice-interaction items
described basic voice performance, including understanding, memory across turns,
response relevance, waiting, turn-taking cues, and repair. The speech--Canvas
items asked how speech and Canvas changes related, complemented each other, and
remained easy to follow as content grew (Appendix~\ref{app:questionnaires}).
For RQ1, we scored SUS and analyzed the study-specific
items individually using descriptive statistics and response distributions.
These results also provided context for interpreting the qualitative
findings.

\textbf{Interviews and Session Reflections.} The approximately 30-minute semi-structured interviews asked about prior
voice-agent use, experiences with \textsc{VoCa}, perceived differences from other
voice agents, unmet expectations, and desired improvements
(Appendix~\ref{app:interview-guide}). Session Reflections were submitted in
\textsc{VoCa} and stored with their Sessions; each comprised a five-point overall
experience rating and one open response to three prompts: whether the
Session helped with the participant's topic, one of four rotating prompts, and
an invitation for further comments
(Appendix~\ref{app:session-reflection}). These accounts informed RQ1 and RQ3 and
supplemented RQ2.
Two researchers independently coded these accounts at the level of meaningful passages, identifying experiences, use contexts,
difficulties, and desired improvements. The researchers compared codes
from both sources within and across participants, discussing differences and grouping
related codes into categories and themes while retaining negative and conflicting
accounts.

\textbf{Interaction logs.} These logs contained each Session's transcript, final Canvas
content, and event history, including user selection and reference,
interruptions, Canvas actions, execution failures, and timing. These logs mainly
informed RQ2 and also supported RQ1. We analyzed them at two levels. At the
Session level, we derived purpose categories.
Full Session transcripts were read and summarized in terms of
purposes and changes of direction. Comparing these summaries across Sessions
produced the purpose categories.
For the descriptive use summary, two researchers manually assigned
applicable purpose categories to every Session, using user requests in
context and, where available, Session-specific reflections. Each
category was counted once per Session and
once per participant; unsuccessful attempts were retained, and Sessions without
an identifiable purpose were reported separately.
At the episode level, episodes spanning different purposes were selected for closer
analysis, including ordinary use, breakdowns, and Sessions without a
reflection; Session ratings helped identify contrasting experiences, such as
higher- and lower-rated Sessions from the same participant. Within
these episodes, user requests, agent speech, Canvas actions, and subsequent responses
were traced and compared, and comparison across episodes identified the ways of interacting. Session-specific reflections and identifiable interview
accounts were linked to the corresponding Sessions; broader accounts informed cross-case interpretation.

\section{Findings}
\label{sec:findings}

Across the five-day deployment, the 18 participants engaged in
251 Sessions, with a median of 15 Sessions per participant (range: 6--21).
These Sessions contained 1,307 submitted turns (median 4 per Session, range: 1--20); 1,088 (83.2\%) used manual
voice and 219 (16.8\%) used real-time voice.
Of the 251 Sessions, 212 (84.5\%) had three or more turns, 23 had ten or
more, and 204 (81.3\%) had a Session Reflection.
We first report participants' assessments of usability and their experiences
of \textsc{VoCa}, followed by what they used
the system for and how they engaged in speech--Canvas interaction with it. We then describe the
improvements participants wanted from \textsc{VoCa}. Appendix~\ref{app:session-examples}
provides example Canvas snapshots from the deployment Sessions discussed below.

\subsection{Usability and Experiences of Speech--Canvas Interaction with \textsc{VoCa} (RQ1)}
\label{sec:findings-rq1}

\begin{figure*}[t]
  \centering
  \includegraphics[width=\textwidth]{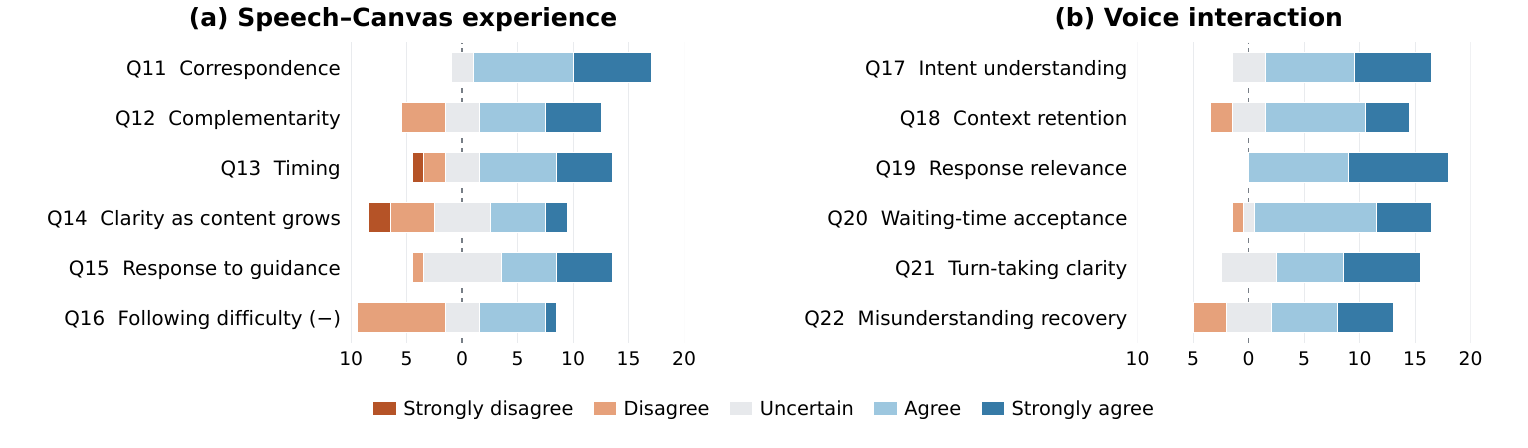}
  \caption{Post-deployment questionnaire responses from 18 participants
  for (a) speech--Canvas experience and (b) voice interaction. Segment widths
  indicate response counts. Disagreement extends left and agreement right;
  the middle response is centered on the dashed line. Item numbers follow
  Appendix~\ref{app:questionnaires}. Each item includes 18 responses.
  Q16 is negatively worded, so agreement indicates greater difficulty
  following both outputs.}
  \label{fig:questionnaire-likert}
  \Description{Two side-by-side diverging stacked bar charts show responses
  to six speech--Canvas experience items and six voice-interaction items.
  The five categories run from strongly disagree to strongly agree, with
  Uncertain in the middle. Each item includes eighteen responses.
  All eighteen respondents agree or strongly agree
  that spoken responses are relevant. For Canvas clarity as content grows,
  seven agree or strongly agree, five fall in the middle category, and six disagree or
  strongly disagree.
  For Q16, eight disagree that following both outputs
  is difficult, three select Uncertain, and seven agree or strongly agree.}
\end{figure*}

Overall, participants found \textsc{VoCa} easy to use and its speech--Canvas
responses helpful and different from those of existing voice agents, but their
experiences varied with how well \textsc{VoCa} chose what and how to show, whether speech and Canvas worked together, and how the Canvas developed.
The mean System Usability Scale score was 76.9
($SD=10.8$), corresponding to ``good''
usability~\cite{bangor2009sus}.
Ratings of the speech--Canvas items were more mixed
(Figure~\ref{fig:questionnaire-likert}a).
On the voice-interaction items (Figure~\ref{fig:questionnaire-likert}b), participants generally rated response relevance, understanding, and waiting time
positively; among completed turns, the median latency from
input submission to logged speech playback onset was 4.42 seconds
(IQR: 3.83--5.43 seconds;
$n=836$ valid turn-level observations).
Ratings of retention across turns, turn-taking cues, and
redirecting after misunderstandings were less consistent.
Technical problems also disrupted some turns: logs explicitly recorded 73 execution
failures (5.6\%~of turns), involving network or service errors, the model returning empty or invalid output or timing out, and Canvas operations that could not be carried out.

\subsubsection{\texorpdfstring{\textbf{How Well Did \textsc{VoCa} Choose What and How to Show?}}{How Well Did VoCa Choose What and How to Show?}}
\label{sec:findings-rq1-fit}

\textbf{Participants valued Visual Objects that organized what they heard for explanation and comparison, making relationships, sequences, and key points intuitive to see.}
When asking how to store different fruits, P03 received a
comparison table without specifying that format:
\emph{``It can directly list a table, with comparisons that show the results clearly. It is the first time I have seen a table drawn on the Canvas.''}
Others agreed: P05, P11, and P12 found the Visual Representations clear, P10 found key points easier to grasp than by listening, and P14 found Visual Objects clearer and more concise than the long answers of other voice agents.

\textbf{What \textsc{VoCa} showed did not always fit users' purposes or provide the detail they needed.} P01 appreciated the grouping and comparison of musical
instruments but noted: \emph{``The way it compared them didn't quite fit what matters
to me when learning an instrument.''} P01 also felt that their reasons and goals
for learning an instrument had not been sufficiently explored. In a technical
explanation, P12 praised the
detailed spoken account but found the words on the diagrams insufficient to understand the
subject; the conversation included repeated requests for more concrete and accessible
explanations. P02 and P18 also described content that was too general: a large table (Discussion Frame) with only general text (P02), and explanations that took up space but helped little (P18).

\textbf{The form \textsc{VoCa} gave the content was hard to predict and not always the one participants wanted.}
P02, P11, and P15 could not tell how \textsc{VoCa} decided which form to use, and P08 expected a mind map and P18 a structure diagram of a model, but found a list of points and mostly text. Forms could also work against reading: P06 found the Canvas too large for answers given in a few points, and P12 found that a mind map (Visual Representation) of tourist attractions ``adds to the reading barrier'' and wished to choose how content was shown.

\subsubsection{\textbf{How Did Participants Follow Speech and Canvas Together?}}
\label{sec:findings-rq1-following}

\textbf{Participants could connect speech with Canvas changes, but the two did not always complement each other or let them keep pace.}
On the speech--Canvas items (Figure~\ref{fig:questionnaire-likert}a), 16 of 18 respondents
agreed or strongly agreed that they understood the relationship between speech and Canvas
changes, and 12 endorsed the timing of those changes; 11 agreed that the two complemented each other, and seven that they sometimes struggled to follow both at once.
Highlights helped connect what was said with what was shown: P02, P05, and P15 valued highlighting of the parts being explained, although timing sometimes slipped, with the Canvas lagging behind speech for P06 and running ahead of it for P19.

\textbf{What was spoken but never reached the Canvas was gone once said, and the Canvas sometimes contradicted speech.} P16, who disagreed that speech and Canvas complemented each other, was frustrated that what the agent said did not appear on the Canvas, and P01 and P14 could not remember or find it again. In one of P18's Sessions, the agent said it would highlight content in yellow, but the highlight appeared in blue.
Even when the two corresponded, participants could not always keep pace with speech or read the relevant Canvas content. The Canvas helped P11 keep up with fast speech, but P02 noted:
\emph{``Sometimes it talks quite fast in long stretches. Even with the Canvas content, it's still a bit hard to follow, so I need to look back.''}
P15 did not know when to interrupt long replies.
Readability also mattered: P19 valued the table format but reported that showing the whole
table left its text too small to read.

\subsubsection{\textbf{How Did Canvas Content Develop as the Conversation Unfolded?}}
\label{sec:findings-rq1-develop}

\textbf{Existing Canvas content could support further questions and additions.} P11 returned to an
earlier menu after asking about wine, and the agent attached further explanations to the
menu. They described it as \emph{``like a topic anchor''}, helping them
see earlier content and resume questions about it. P04 found that mind maps in later turns connected well with the earlier one, although each turn's map remained separate.

\textbf{The agent took up later requests, but the Canvas did not always develop with them.}
Ten of 18 respondents agreed or strongly agreed that the agent adjusted
subsequent responses when guided or corrected.
P16 considered asking for changes acceptable, explaining:
\emph{``It understands what you want to change, but the result might not match what
you have in mind.''}
P14 recalled clear summaries and appealing charts
for a short-story collection, but when they requested further analysis of the characters'
relationships, the responses became repetitive and the charts remained in their initial
form. They reported turning to a search engine to continue the task. P03 and P19 also reported that the Canvas stopped updating after the first turn or two.

\textbf{Participants differed on whether the Canvas remained clear in Sessions with substantial content.}
Seven~of 18 respondents agreed and six disagreed.
While developing a swimming plan, P15 asked for a tutorial-link column in the existing
stage table. The agent left the tutorials in a separate table and annotated how the
material should be combined, but did not add the requested column to the stage table. They appreciated the initial plan and
annotations, yet reported that subsequent tables increased their cognitive burden. By
contrast, P03 praised a new medication-comparison table linked to an earlier overview for
making the follow-up's connection clear.

\subsection{\texorpdfstring{How Participants Used \textsc{VoCa} (RQ2)}{How Participants Used VoCa (RQ2)}}
\label{sec:findings-rq2}

\subsubsection{\texorpdfstring{Purposes of Use}{Purposes of Use}}

Table~\ref{tab:everyday-uses} summarizes six purposes of use identified through
Session coding. \textbf{Learning and understanding, decision-making and planning, and
information seeking and practical help appeared most often, each in Sessions from at least 15 of the 18 participants.} Participants
also used \textsc{VoCa} for emotional support, entertainment and casual
conversation, and creative work.
These purposes could overlap within a Session as participants combined
activities or changed direction during the conversation; 23.9\% (60) of
Sessions had two or more purposes, most often information seeking together with
learning or planning.
Outside the table, 13.9\% (35) of Sessions involved exploring the system's
capabilities (18 of them alongside a listed purpose), and 3.2\% (8) had no
identifiable purpose.

\begin{table*}[t]
  \caption{Purposes of use identified through Session coding, ordered by number
  of Sessions. Values are percentage (count) of 251 Sessions. Purposes
  could overlap within a Session.
  Counts include unsuccessful attempts and Sessions that also involved
  system exploration; they describe recorded use, not task success.}
  \label{tab:everyday-uses}
  \centering
  \small
  \setlength{\tabcolsep}{5pt}
  \renewcommand{\arraystretch}{1.15}
  \begin{tabular}{@{}>{\raggedright\arraybackslash}p{0.21\textwidth}
                    >{\raggedright\arraybackslash}p{\dimexpr0.67\textwidth-20pt\relax}
                    >{\raggedleft\arraybackslash}p{0.12\textwidth}@{}}
    \toprule
    \leavevmode Purpose & \leavevmode Description & Sessions \\
    \midrule
    Learning and understanding &
    Participants asked successive questions on topics such as mathematical proofs, language rules, philosophical ideas, and AI concepts; some also practiced speaking English or worked through example problems. The agent often drew concepts and their relations as diagrams, and participants sometimes selected earlier content when following up. &
    33.1\% (83) \\
    \addlinespace
    Decision-making and planning &
    \leavevmode Participants made decisions, for example about gifts, travel, or study, weighing options against their preferences, budgets, or time, or developed plans such as schedules and task lists. Options and plans often appeared in Discussion Frames and were updated in later turns; Badges marked recommendations, cautions, or completed tasks. &
    33.1\% (83) \\
    \addlinespace
    \leavevmode Information seeking and practical help &
    Participants looked up information, for example on weather or products, or sought practical help with a task at hand, such as finding a Photoshop function or dealing with food sticking to a pan. Steps, settings, or facts were mostly listed in Notes, and Annotations were often added in later turns. &
    28.7\% (72) \\
    \addlinespace
    Emotional support &
    Participants talked through difficulties such as anxiety, rejection, loneliness, or pressures around study or work, seeking reassurance, help making sense of their feelings, or companionship. Notes or diagrams set out their situation or ways to cope, and Annotations added advice, in some Sessions including when to seek professional help. &
    8.0\% (20) \\
    \addlinespace
    Entertainment and casual conversation &
    \leavevmode Participants engaged in activities such as chatting about favorite songs, role-playing with fictional personas, fortune-telling, and song-elimination games. Discussion Frames often tracked game rounds, tarot draws, or choices, and Notes held role-play lines and lists. &
    6.8\% (17) \\
    \addlinespace
    \leavevmode Ideation and creative work &
    Participants worked on ideas and writing, such as developing product ideas, drafting and revising emails or stories, or working out how to make music or videos with AI. Written drafts appeared in Notes, with revised versions added on request, and Discussion Frames laid out workflows, in one Session dividing music-making between the participant and AI. &
    6.0\% (15) \\
    \bottomrule
  \end{tabular}
\end{table*}

\subsubsection{\texorpdfstring{Speech--Canvas Interaction in Use}{Speech--Canvas Interaction in Use}}
\label{sec:findings-rq2-interaction}

Participants used speech and the Canvas together in different ways, from how they followed a single response to how they continued a discussion across breaks.

\textbf{Participants did not always listen and look at the same time.} P08 described
putting their phone aside during a response, then consulting its visual summary:
\emph{``I might put my phone aside, and after it finishes responding,
I take a look at the mind map.''}
They described listening and looking as staggered, and at times skipped the speech to go straight to the key points on the Canvas. P09 valued being able to stop a long-winded reply and turn to the Canvas. For these participants, the Canvas was less something to follow during speech than a summary to consult (Section~\ref{sec:findings-rq1-following}).

\textbf{Participants asked the agent to write down what they wanted to keep.}
While learning Dutch word order, P04 asked the agent
to write down its examples:
\emph{``My Dutch is poor; I cannot remember it just from hearing you say it.''}
The agent created a Note containing examples and meanings, then added English
equivalents as Annotations when P04 requested them
(Figure~\ref{fig:session-dutch}). Likewise, P02 found the spoken
Photoshop guidance too extensive and fast, and requested a lookup list of
functions and menu locations. P01 asked for a to-do list with a checkbox on the left and a line of text on the right, and P09 asked the agent to write out an email to a teacher on the Canvas. P04 and P16 also asked for spoken content missing from the Canvas to be added.

\textbf{Participants also read what was on the Canvas and built further questions and additions on it.}
In 37 Sessions from 14 participants, a turn was submitted while an object on the Canvas was selected.
In a philosophy discussion, P16 asked what two arrows pointing toward
existentialism meant in an earlier Visual Representation, and the agent explained
the relationships through speech and added an Annotation to the diagram
(Figure~\ref{fig:session-philosophy}).
P02 had the agent mark unbought items on a Note of dormitory supplies, then asked which small items could be stored together.
Similarly, P04 referred to entries in an album timeline to identify
errors and omissions and requested revisions.

\textbf{This use of existing material also extended across breaks in the conversation.}
P05, who had dictated a task list intending to keep the Session open for several days, returned about 35 hours later to report two completed tasks, and the agent added completion Badges to the original entries (Figure~\ref{fig:session-task-updates}). P10 came back three hours later to ask about side dishes for a meal discussed earlier, and the agent annotated the earlier Note. This did not always work: P16 returned after about eight hours to have a music workflow reordered, but the response was interrupted and the Frame stayed unchanged.

\subsection{\texorpdfstring{Desired Improvements for \textsc{VoCa} (RQ3)}{Desired Improvements for VoCa (RQ3)}}
\label{sec:findings-rq3}

Participants wanted to better understand and steer how the conversation with \textsc{VoCa} unfolded, and to connect it with their own materials and with later work.

\subsubsection{\texorpdfstring{Knowing What \textsc{VoCa} Can and Will Do}{Knowing What VoCa Can and Will Do}}
\label{sec:findings-rq3-know}

\textbf{With the agent leading what appeared on the Canvas, participants wanted to
know more about what \textsc{VoCa} could do and how they could take part.} P05 wanted
the Canvas to introduce \textsc{VoCa}'s functions when asked, and P15 wanted clearer
indications of whether and how they could adjust what it produced. P03 wanted to
circle part of the Canvas to ask about it, not knowing that selecting an object
already allowed this.

\textbf{Participants also wanted the agent to be open about what it could not do
and about failures.} P09 wanted it to say truthfully what it could do, after it first
claimed it could do anything, and P17 tried eight times to have it draw a circuit board
without being told that this was beyond its abilities. P02 wanted clearer messages when
Canvas generation failed or a response came back empty, and P14 wanted to correct
misrecognized input text.

\textbf{Participants also wanted capabilities of the underlying model that the
prototype did not provide.} Seven participants noted that the agent could not search
the web for current information, and P02, P04, and P07 asked for this. P12 wanted help
with complex questions that required longer reasoning, which sometimes produced no
output.

\subsubsection{More Flexible Ways to Modify Canvas Content}
\label{sec:findings-rq3-modify}

\textbf{Requests for direct Canvas interaction ranged from rearranging existing
content to adding users' own writing and drawings.} The prototype did not support
direct editing or fine-grained layout adjustment by users. Twelve of the 18 participants nevertheless wanted to do so, for example to move overlapping text or to edit or delete the agent's content. After encountering overlapping
information, P12 commented:
\emph{``I'd like to drag the overlapping content apart myself.''}
P10 went further, proposing that users could write, draw, or modify
content on the Canvas and that the agent could recognize these contributions
and respond to them.

\textbf{For revisions made through speech, participants expected the agent to
modify the Canvas content already in use.} P18 wanted to add branches within an existing process
diagram through spoken requests, explaining that a separate, connected diagram
did not incorporate the branches into the original flow.
P04 and P12 likewise wanted later changes made on the original mind map rather than in separate maps or added notes (Section~\ref{sec:findings-rq1-develop}), P16 wanted local edits rather than regenerating content, which introduced errors, and P05 wanted to delete an earlier flowchart version once a clearer one was generated.
At the same time, P18 valued the visible strike-throughs left when a timetable
was revised.
The amount of manual correction mattered too: P03 did not want to
fix numerous mathematical errors by hand.

\subsubsection{\texorpdfstring{Adapting \textsc{VoCa} to Users' Goals and Pace}{Adapting VoCa to Users' Goals and Pace}}
\label{sec:findings-rq3-adapt}

\textbf{What participants expected from the agent depended on what they were
trying to do.} When planning a schedule, P01 wanted the agent to ask about their
fixed commitments after outlining the plan's sections. They had sought help
thinking through the plan and did not know how to continue when the agent stopped
at the outline.
P07 and P15 likewise wanted the agent to ask about missing information before proceeding.
In a Session Reflection, P15 was dissatisfied with the agent treating
emotional remarks as problems to solve and explained:
\emph{``I hope it can tell whether I just want to chat and explore ideas
or I'm really trying to get something done.''}

\textbf{Participants also asked for ways to adjust how content appeared and
how they took part in the conversation.}
P01 sometimes had a diagram or table in mind
but did not usually specify that form when making a request, and suggested
templates as a way to express this preference. They also acknowledged that a
different form chosen by the agent could be reasonable.
P15 likewise wanted to choose from templates, such as a study-plan format that could be copied into a document.
Other requests concerned the pace of the conversation. P03 wanted Canvas content
to appear gradually alongside the spoken explanation, rather than all at once.
P19 wanted the key points of long spoken replies written down so that they would not lose track of their own questions, and P06 and P08 preferred manual voice.
For P02, having time to formulate a response mattered: they asked to retain
manual control over starting and ending recording because automatic input
left too little time to think in that situation.

\subsubsection{Using Images, Documents, and More Visual Forms with \textsc{VoCa}}
\label{sec:findings-rq3-media}

\textbf{Participants wanted to share their own images and documents with the agent
and to see more kinds of visuals on the Canvas.} P02, P05, and P14 wanted to upload
images or files for the agent to work with, which P02 valued more than image
generation. P03, P06, P09, and P11 wanted images on the Canvas, such as a textbook-like
illustration in place of a flowchart-like diagram, and P18 wanted Canvas elements
themselves to move and change, such as data passing through a model.

\subsubsection{Using Content from \textsc{VoCa} in Subsequent Activities}

\textbf{Beyond the conversation itself, participants saw potential uses for
\textsc{VoCa}'s content in further design work and everyday activities.}
Some wanted to continue the work in another tool. P07 hoped to take a poster
plan into Figma to continue designing and make precise adjustments.
P05 wanted to organize ideas with \textsc{VoCa} and then pass the content to a
coding agent to carry out the work.
For P08, the desired outcome was a list to save or print and keep on the desk:
\emph{``But I need something I can actually use,''} they explained.
These wishes continue the requests, made during conversations, to have the agent write down what participants wanted to keep (Section~\ref{sec:findings-rq2-interaction}).

\section{Discussion}
\label{sec:discussion}

\subsection{\texorpdfstring{Voice Agents with Generative Visual Interfaces}{Voice Agents with Generative Visual Interfaces}}
\label{sec:discussion-interaction}

Our deployment suggests that a voice agent that generates visual
content as it speaks can help people understand content, express their needs,
and develop ideas with the agent. This value reflects the practice that
motivated \textsc{VoCa}: people talking at a whiteboard decide as they speak
what to write, draw, or point to, and the board develops with their discussion.
Participants valued forms the agent chose and used visible content to look back
and build on across turns and breaks, in line with prior work in which
visuals supported spoken
communication~\cite{liu2023visualcaptions,ge2025gencomui}. Our study examines this
potential across open-ended conversations.
There, generated visual content can serve as a medium between the user
and the agent that moves the conversation forward, not bound to the order in
which things were said: it
gives users something to question, clarify, or build on, while
speech lets them explain what matters and redirect the agent.

Prior work has used generated visual forms to support human--AI
communication, as in Graphologue, Jelly, and
Orality~\cite{jiang2023graphologue,cao2025malleable,li2026orality}.
\textsc{VoCa} offers a concrete exploration of how a voice agent can
coordinate its speech with a generative visual interface: it speaks while creating, pointing to,
annotating, and developing content on the Canvas. The design challenge is to connect
these actions with the unfolding conversation: visually clear content can still
miss the user's purpose, and an appropriate explanation can reveal a need to
change what is shown. Our results showed both: forms the agent chose
did not always fit what users needed, and requests to restructure content in
use were not met. These reflect design choices in \textsc{VoCa}, such as an
agent that acts on its own initiative and appends rather than revises content,
and other choices remain to be explored. Our design space organizes these choices across expression,
actions, timing, and participation; our results illustrate why they need to
be considered together.

\subsection{Application Opportunities}
\label{sec:discussion-opportunities}

How participants used \textsc{VoCa} suggests where voice agents with generative
visual interfaces could help: in learning, work, and daily life. When people learn,
an agent could explain a topic using different visual forms, much like a human
teacher. At work, it could act as a partner with whom people discuss, plan, and make
decisions on a shared Canvas. In daily life, it could help people organize their
thoughts and make plans. Beyond the personal digital devices people use daily, such
an agent could also take other forms. For example, a robot could make full use of
its screen~\cite{kim2026speaking} when communicating with people, showing and
developing content as it speaks. On a meeting-room whiteboard, the setting introduced in
our formative study, an agent could act as a secretary for people working together,
organizing ideas, decisions, and action items while
contributing summaries or suggestions.

For such uses, a voice agent would also need to fit into the tools people
already use: participants wanted the capabilities they knew from other AI tools
and ways to carry content into their existing workflows. For example, integrated
into platforms such as Claude Code or Codex, it could let users discuss ideas
whenever and wherever they arise and then pass them to these tools for implementation.

\subsection{Design Implications}
\label{sec:discussion-implications}

We propose four design considerations for voice agents that coordinate speech
and Canvas actions, grounded in our design exploration and user study.

\subsubsection{Combine Agent Assistance with User Contributions}

\textbf{Users should be able to contribute directly to the Canvas while retaining
the agent's initiative in developing content.} \textsc{VoCa} left Canvas
work to the agent, with users selecting, navigating, and moving whole objects
(D4), yet users also wanted to write, draw, and edit. Designers could allow these contributions to guide the
agent's subsequent speech and Canvas actions. For example, a user might sketch
an initial idea, which the agent could build on by suggesting additions and
developing the proposal through conversation.

\subsubsection{Adapt Speech and Canvas Actions to Conversational Intentions}

\textbf{An agent should adapt how it uses the Canvas to what the user is
trying to do, rather than using it the same way in every conversation.}
Users' expectations differed between exploring ideas and working toward
a result, so the agent should judge this intention from the conversation as a
whole. For a result, such as a plan, it might ask about constraints and extend a visible plan; for
open-ended talk, it might explore
an idea through questions without creating new Canvas content. When the user's
intentions are unclear, the agent could clarify them before deciding how to
respond, or propose a Canvas action for the user to accept instead of
acting on its own initiative (D6).

\subsubsection{Decide What to Say, What to Show, and When}

\textbf{Agents should coordinate how much they say, what they leave
visible, and the pace at which they introduce both, adjusting them
as the conversation unfolds.} Designing multimodal responses involves selecting forms of expression and
coordinating their timing~\cite{ma2019design,ge2025gencomui}; our
design space treats these as content (D1) and temporal coordination (D5), but an
agent should choose them together. Like a person at a whiteboard, it could judge
what to write down and when to slow down from signals such as gaze (D4) that
show whether the user is keeping up; \textsc{VoCa} used no such signals, and some
users struggled to follow speech and Canvas at once. For example, an agent could
keep an example and its key terms visible while explaining their relationships
in shorter spoken segments.

\subsubsection{Distinguish Adding Information from Revising Existing Content}

\textbf{Canvas changes should match whether a request extends the record
of the conversation or revises the material in use.}
Our design space distinguishes appending, revising, and versioning content
(D2). Revising in place suits a structure users are still working with, such as
a table that needs another column, whereas appending or keeping versions with
revision marks suits content whose development users want to follow.

\section{Limitations and Future Work}
\label{sec:limitations}

\subsection{Limitations}
\label{sec:limitations-scope}

Our findings should be read in light of three limitations. First, the
version of \textsc{VoCa} evaluated in this study represents one way of
designing speech--Canvas interaction: the agent was the principal Canvas actor,
revisions were expressed through additions and Annotations, and users did not
directly edit content or make fine-grained layout adjustments. We therefore
cannot separate whether participants' requests to edit and add to Canvas content
arose from these limits, from expectations formed with manipulable visual
interfaces, or from their own need to express ideas visually. Although
\textsc{VoCa} was designed for real-time conversation, its speech and Canvas
actions came with noticeable latency: speech began a median of 4.42 seconds after
input~\cite{maslych2025mitigating}. Participants' experiences may also have
been influenced by the prototype's current visual repertoire and
occasional issues with generation,
speech recognition, readability, and recovery.

Second, the sample was small and consisted of university students or recent
graduates, and people with other backgrounds may use such agents
differently.

Finally, the study was a five-day deployment rather than a controlled
comparison, which limits what its data can show. The findings
therefore do not identify the effects of individual design choices or comparative
advantages over speech-only agents. The interaction logs captured speech
and Canvas actions but not gaze or comprehension, so they cannot show in
detail how the Canvas shaped the conversation. Accordingly, we use these data
to inform design rather than make causal claims.

\subsection{Future Work}
\label{sec:future-work}

Future work could address these limitations. Future versions of
\textsc{VoCa} could let users edit and arrange Canvas content directly, and
studying how people use this control could show whether participants' requests
arose from \textsc{VoCa}'s limits or from their own need to express ideas
visually. They could also reduce latency and broaden the visual forms available
to the agent. Larger and more varied samples could show how people beyond
university students use such agents. Finally, controlled studies could compare
\textsc{VoCa} with speech-only agents and vary individual design choices, and
longer deployments could show how use develops over time. Combining eye
tracking with time-aligned action logs could show when users look at referenced
content or return to earlier parts of the Canvas, while retrospective interviews and
comprehension measures would help interpret how the Canvas shapes the
conversation.

\section{Conclusion}
\label{sec:conclusion}

This work explored the value of speech--canvas interaction between people and
voice agents. Through a formative study, we developed a design space and
informed the design of \textsc{VoCa}, a voice agent that coordinates speech and
canvas actions. Our exploratory deployment illustrated how participants used
this interaction in learning, work, and daily life. Their experiences suggest that
a voice agent that generates visual content as it speaks can help people
understand content, express their needs, and develop ideas with it. Realizing
this value requires content in forms suited to the conversation, speech
and canvas actions that users can follow and influence, and content that develops
as the conversation unfolds. Together, these contributions inform the design of
voice agents with generative visual interfaces.

\bibliographystyle{ACM-Reference-Format}
\bibliography{references}


\begin{thebibliography}{62}


\ifx \showCODEN    \undefined \def \showCODEN     #1{\unskip}     \fi
\ifx \showISBNx    \undefined \def \showISBNx     #1{\unskip}     \fi
\ifx \showISBNxiii \undefined \def \showISBNxiii  #1{\unskip}     \fi
\ifx \showISSN     \undefined \def \showISSN      #1{\unskip}     \fi
\ifx \showLCCN     \undefined \def \showLCCN      #1{\unskip}     \fi
\ifx \shownote     \undefined \def \shownote      #1{#1}          \fi
\ifx \showarticletitle \undefined \def \showarticletitle #1{#1}   \fi
\ifx \showURL      \undefined \def \showURL       {\relax}        \fi
\providecommand\bibfield[2]{#2}
\providecommand\bibinfo[2]{#2}
\providecommand\natexlab[1]{#1}
\providecommand\showeprint[2][]{arXiv:#2}

\bibitem[Ainsworth(2006)]%
        {ainsworth2006deft}
\bibfield{author}{\bibinfo{person}{Shaaron Ainsworth}.}
  \bibinfo{year}{2006}\natexlab{}.
\newblock \showarticletitle{{DeFT}: A Conceptual Framework for Considering
  Learning with Multiple Representations}.
\newblock \bibinfo{journal}{\emph{Learning and Instruction}}
  \bibinfo{volume}{16}, \bibinfo{number}{3} (\bibinfo{year}{2006}),
  \bibinfo{pages}{183--198}.
\newblock
\href{https://doi.org/10.1016/j.learninstruc.2006.03.001}{doi:\nolinkurl{10.1016/j.learninstruc.2006.03.001}}


\bibitem[Amin et~al\mbox{.}(2026)]%
        {amin2026conversations}
\bibfield{author}{\bibinfo{person}{Rifat~Mehreen Amin},
  \bibinfo{person}{Alperen Adatepe}, \bibinfo{person}{Daniela Fernandes},
  \bibinfo{person}{Daniel Buschek}, {and} \bibinfo{person}{Andreas Butz}.}
  \bibinfo{year}{2026}\natexlab{}.
\newblock \bibinfo{title}{Conversations in Space: Structuring Non-Linear {LLM}
  Interactions on a Canvas}.
\newblock
\showeprint[arxiv]{2605.15848}~[cs.HC]
\href{https://doi.org/10.48550/arXiv.2605.15848}{doi:\nolinkurl{10.48550/arXiv.2605.15848}}


\bibitem[Ammari et~al\mbox{.}(2019)]%
        {ammari2019music}
\bibfield{author}{\bibinfo{person}{Tawfiq Ammari}, \bibinfo{person}{Jofish
  Kaye}, \bibinfo{person}{Janice~Y. Tsai}, {and} \bibinfo{person}{Frank
  Bentley}.} \bibinfo{year}{2019}\natexlab{}.
\newblock \showarticletitle{Music, Search, and {IoT}: How People (Really) Use
  Voice Assistants}.
\newblock \bibinfo{journal}{\emph{ACM Transactions on Computer-Human
  Interaction}} \bibinfo{volume}{26}, \bibinfo{number}{3}, Article
  \bibinfo{articleno}{17} (\bibinfo{year}{2019}), \bibinfo{numpages}{28}~pages.
\newblock
\href{https://doi.org/10.1145/3311956}{doi:\nolinkurl{10.1145/3311956}}


\bibitem[Baghdadi et~al\mbox{.}(2025)]%
        {baghdadi2025hud}
\bibfield{author}{\bibinfo{person}{Mahmoud Baghdadi}, \bibinfo{person}{Dilara
  Samad-Zada}, {and} \bibinfo{person}{Achim Ebert}.}
  \bibinfo{year}{2025}\natexlab{}.
\newblock \showarticletitle{Evaluating Rich Visual Feedback on Head-Up Displays
  for In-Vehicle Voice Assistants: A User Study}.
\newblock \bibinfo{journal}{\emph{Multimodal Technologies and Interaction}}
  \bibinfo{volume}{9}, \bibinfo{number}{11}, Article \bibinfo{articleno}{114}
  (\bibinfo{year}{2025}).
\newblock
\href{https://doi.org/10.3390/mti9110114}{doi:\nolinkurl{10.3390/mti9110114}}


\bibitem[Bangor et~al\mbox{.}(2009)]%
        {bangor2009sus}
\bibfield{author}{\bibinfo{person}{Aaron Bangor}, \bibinfo{person}{Philip
  Kortum}, {and} \bibinfo{person}{James Miller}.}
  \bibinfo{year}{2009}\natexlab{}.
\newblock \showarticletitle{Determining What Individual {SUS} Scores Mean:
  Adding an Adjective Rating Scale}.
\newblock \bibinfo{journal}{\emph{Journal of Usability Studies}}
  \bibinfo{volume}{4}, \bibinfo{number}{3} (\bibinfo{year}{2009}),
  \bibinfo{pages}{114--123}.
\newblock


\bibitem[Brooke(1996)]%
        {brooke1996sus}
\bibfield{author}{\bibinfo{person}{John Brooke}.}
  \bibinfo{year}{1996}\natexlab{}.
\newblock \showarticletitle{{SUS}: A Quick and Dirty Usability Scale}.
\newblock In \bibinfo{booktitle}{\emph{Usability Evaluation in Industry}},
  \bibfield{editor}{\bibinfo{person}{Patrick~W. Jordan}, \bibinfo{person}{Bruce
  Thomas}, \bibinfo{person}{Ian~L. McClelland}, {and} \bibinfo{person}{Bernard
  Weerdmeester}} (Eds.). \bibinfo{publisher}{Taylor \& Francis},
  \bibinfo{address}{London, UK}, \bibinfo{pages}{189--194}.
\newblock


\bibitem[Cao et~al\mbox{.}(2025)]%
        {cao2025malleable}
\bibfield{author}{\bibinfo{person}{Yining Cao}, \bibinfo{person}{Peiling
  Jiang}, {and} \bibinfo{person}{Haijun Xia}.} \bibinfo{year}{2025}\natexlab{}.
\newblock \showarticletitle{Generative and Malleable User Interfaces with
  Generative and Evolving Task-Driven Data Model}. In
  \bibinfo{booktitle}{\emph{Proceedings of the 2025 CHI Conference on Human
  Factors in Computing Systems}}. \bibinfo{publisher}{ACM},
  \bibinfo{pages}{1--20}.
\newblock
\href{https://doi.org/10.1145/3706598.3713285}{doi:\nolinkurl{10.1145/3706598.3713285}}


\bibitem[Chen et~al\mbox{.}(2023)]%
        {chen2023screen}
\bibfield{author}{\bibinfo{person}{Chen Chen}, \bibinfo{person}{Ella~T.
  Lifset}, \bibinfo{person}{Yichen Han}, \bibinfo{person}{Arkajyoti Roy},
  \bibinfo{person}{Michael Hogarth}, \bibinfo{person}{Alison~A. Moore},
  \bibinfo{person}{Emilia Farcas}, {and} \bibinfo{person}{Nadir Weibel}.}
  \bibinfo{year}{2023}\natexlab{}.
\newblock \showarticletitle{Screen or No Screen? Lessons Learnt from a
  Real-World Deployment Study of Using Voice Assistants With and Without
  Touchscreen for Older Adults}. In \bibinfo{booktitle}{\emph{Proceedings of
  the 25th International ACM SIGACCESS Conference on Computers and
  Accessibility}}. \bibinfo{publisher}{ACM}, \bibinfo{pages}{1--21}.
\newblock
\href{https://doi.org/10.1145/3597638.3608378}{doi:\nolinkurl{10.1145/3597638.3608378}}


\bibitem[Chen et~al\mbox{.}(2026)]%
        {chen2026generative}
\bibfield{author}{\bibinfo{person}{Jiaqi Chen}, \bibinfo{person}{Yanzhe Zhang},
  \bibinfo{person}{Yutong Zhang}, \bibinfo{person}{Yijia Shao}, {and}
  \bibinfo{person}{Diyi Yang}.} \bibinfo{year}{2026}\natexlab{}.
\newblock \showarticletitle{Generative Interfaces for Language Models}. In
  \bibinfo{booktitle}{\emph{Findings of the Association for Computational
  Linguistics: ACL 2026}}. \bibinfo{publisher}{Association for Computational
  Linguistics}, \bibinfo{pages}{1499--1519}.
\newblock
\href{https://doi.org/10.18653/v1/2026.findings-acl.74}{doi:\nolinkurl{10.18653/v1/2026.findings-acl.74}}


\bibitem[Coscia et~al\mbox{.}(2025)]%
        {coscia2025ongoal}
\bibfield{author}{\bibinfo{person}{Adam~J. Coscia}, \bibinfo{person}{Shunan
  Guo}, \bibinfo{person}{Eunyee Koh}, {and} \bibinfo{person}{Alex Endert}.}
  \bibinfo{year}{2025}\natexlab{}.
\newblock \showarticletitle{{OnGoal}: Tracking and Visualizing Conversational
  Goals in Multi-Turn Dialogue with Large Language Models}. In
  \bibinfo{booktitle}{\emph{Proceedings of the 38th Annual ACM Symposium on
  User Interface Software and Technology}}. \bibinfo{publisher}{ACM},
  \bibinfo{pages}{1--18}.
\newblock
\href{https://doi.org/10.1145/3746059.3747746}{doi:\nolinkurl{10.1145/3746059.3747746}}


\bibitem[Dillenbourg and Traum(2006)]%
        {dillenbourg2006sharing}
\bibfield{author}{\bibinfo{person}{Pierre Dillenbourg} {and}
  \bibinfo{person}{David Traum}.} \bibinfo{year}{2006}\natexlab{}.
\newblock \showarticletitle{Sharing Solutions: Persistence and Grounding in
  Multimodal Collaborative Problem Solving}.
\newblock \bibinfo{journal}{\emph{The Journal of the Learning Sciences}}
  \bibinfo{volume}{15}, \bibinfo{number}{1} (\bibinfo{year}{2006}),
  \bibinfo{pages}{121--151}.
\newblock
\href{https://doi.org/10.1207/s15327809jls1501_9}{doi:\nolinkurl{10.1207/s15327809jls1501_9}}


\bibitem[Ellawela et~al\mbox{.}(2025)]%
        {ellawela2025whiteboards}
\bibfield{author}{\bibinfo{person}{Suveen Ellawela}, \bibinfo{person}{Sashenka
  Gamage}, {and} \bibinfo{person}{Dinithi Dissanayake}.}
  \bibinfo{year}{2025}\natexlab{}.
\newblock \bibinfo{title}{Proactive Agentic Whiteboards: Enhancing Diagrammatic
  Learning}.
\newblock
\showeprint[arxiv]{2512.01234}~[cs.HC]
\href{https://doi.org/10.48550/arXiv.2512.01234}{doi:\nolinkurl{10.48550/arXiv.2512.01234}}


\bibitem[Ge et~al\mbox{.}(2024)]%
        {ge2024cocobo}
\bibfield{author}{\bibinfo{person}{Yate Ge}, \bibinfo{person}{Yi Dai},
  \bibinfo{person}{Run Shan}, \bibinfo{person}{Kechun Li},
  \bibinfo{person}{Yuanda Hu}, {and} \bibinfo{person}{Xiaohua Sun}.}
  \bibinfo{year}{2024}\natexlab{}.
\newblock \showarticletitle{{Cocobo}: Exploring Large Language Models as the
  Engine for End-User Robot Programming}. In \bibinfo{booktitle}{\emph{2024
  IEEE Symposium on Visual Languages and Human-Centric Computing}}.
  \bibinfo{publisher}{IEEE}, \bibinfo{pages}{89--95}.
\newblock
\href{https://doi.org/10.1109/VL/HCC60511.2024.00020}{doi:\nolinkurl{10.1109/VL/HCC60511.2024.00020}}


\bibitem[Ge et~al\mbox{.}(2025)]%
        {ge2025gencomui}
\bibfield{author}{\bibinfo{person}{Yate Ge}, \bibinfo{person}{Meiying Li},
  \bibinfo{person}{Xipeng Huang}, \bibinfo{person}{Yuanda Hu},
  \bibinfo{person}{Qi Wang}, \bibinfo{person}{Xiaohua Sun}, {and}
  \bibinfo{person}{Weiwei Guo}.} \bibinfo{year}{2025}\natexlab{}.
\newblock \showarticletitle{{GenComUI}: Exploring Generative Visual Aids as
  Medium to Support Task-Oriented Human-Robot Communication}. In
  \bibinfo{booktitle}{\emph{Proceedings of the 2025 CHI Conference on Human
  Factors in Computing Systems}}. \bibinfo{publisher}{ACM},
  \bibinfo{pages}{1--21}.
\newblock
\href{https://doi.org/10.1145/3706598.3714238}{doi:\nolinkurl{10.1145/3706598.3714238}}


\bibitem[Gergle et~al\mbox{.}(2013)]%
        {gergle2013visual}
\bibfield{author}{\bibinfo{person}{Darren Gergle}, \bibinfo{person}{Robert~E.
  Kraut}, {and} \bibinfo{person}{Susan~R. Fussell}.}
  \bibinfo{year}{2013}\natexlab{}.
\newblock \showarticletitle{Using Visual Information for Grounding and
  Awareness in Collaborative Tasks}.
\newblock \bibinfo{journal}{\emph{Human--Computer Interaction}}
  \bibinfo{volume}{28}, \bibinfo{number}{1} (\bibinfo{year}{2013}),
  \bibinfo{pages}{1--39}.
\newblock
\href{https://doi.org/10.1080/07370024.2012.678246}{doi:\nolinkurl{10.1080/07370024.2012.678246}}


\bibitem[Gmeiner et~al\mbox{.}(2026)]%
        {gmeiner2026pointaloud}
\bibfield{author}{\bibinfo{person}{Frederic Gmeiner}, \bibinfo{person}{John
  Thompson}, \bibinfo{person}{George Fitzmaurice}, {and}
  \bibinfo{person}{Justin Matejka}.} \bibinfo{year}{2026}\natexlab{}.
\newblock \showarticletitle{{PointAloud}: An Interaction Suite for
  {AI}-Supported Pointer-Centric Think-Aloud Computing}. In
  \bibinfo{booktitle}{\emph{Proceedings of the 2026 CHI Conference on Human
  Factors in Computing Systems}}. \bibinfo{publisher}{ACM},
  \bibinfo{pages}{1--37}.
\newblock
\href{https://doi.org/10.1145/3772318.3790797}{doi:\nolinkurl{10.1145/3772318.3790797}}


\bibitem[Ho et~al\mbox{.}(2025)]%
        {ho2025art}
\bibfield{author}{\bibinfo{person}{Hoang~Phuoc Ho}, \bibinfo{person}{Vani
  Ramesh}, \bibinfo{person}{Ivo Zaloudek}, \bibinfo{person}{Delaram~Javdani
  Rikhtehgar}, {and} \bibinfo{person}{Shenghui Wang}.}
  \bibinfo{year}{2025}\natexlab{}.
\newblock \showarticletitle{Enhancing Visitor Engagement in Interactive Art
  Exhibitions with Visual-Enhanced Conversational Agents}. In
  \bibinfo{booktitle}{\emph{Proceedings of the 30th International Conference on
  Intelligent User Interfaces}}. \bibinfo{publisher}{ACM},
  \bibinfo{pages}{660--671}.
\newblock
\href{https://doi.org/10.1145/3708359.3712145}{doi:\nolinkurl{10.1145/3708359.3712145}}


\bibitem[Hoefer et~al\mbox{.}(2025)]%
        {hoefer2025telltime}
\bibfield{author}{\bibinfo{person}{Michael~J. Hoefer}, \bibinfo{person}{Raegan
  Rychecky}, \bibinfo{person}{Max Gong}, {and} \bibinfo{person}{Stephen
  Voida}.} \bibinfo{year}{2025}\natexlab{}.
\newblock \showarticletitle{{TellTime}: An {AI}-Augmented Calendar with a Voice
  Interface for Collecting Time-Use Data}. In
  \bibinfo{booktitle}{\emph{Proceedings of the 30th International Conference on
  Intelligent User Interfaces}}. \bibinfo{publisher}{ACM},
  \bibinfo{pages}{1366--1380}.
\newblock
\href{https://doi.org/10.1145/3708359.3712116}{doi:\nolinkurl{10.1145/3708359.3712116}}


\bibitem[Hombeck et~al\mbox{.}(2026)]%
        {hombeck2026vocalvis}
\bibfield{author}{\bibinfo{person}{Jan Hombeck}, \bibinfo{person}{Henrik
  Voigt}, \bibinfo{person}{Jasna Nuhic}, \bibinfo{person}{Monique Meuschke},
  {and} \bibinfo{person}{Kai Lawonn}.} \bibinfo{year}{2026}\natexlab{}.
\newblock \showarticletitle{{VocalVis}: Bridging Voice-Based Control and 3D
  Surface Visualization}.
\newblock \bibinfo{journal}{\emph{Computers \& Graphics}}
  \bibinfo{volume}{138} (\bibinfo{year}{2026}), \bibinfo{pages}{104632}.
\newblock
\href{https://doi.org/10.1016/j.cag.2026.104632}{doi:\nolinkurl{10.1016/j.cag.2026.104632}}


\bibitem[Huang et~al\mbox{.}(2025)]%
        {huang2025sketchgpt}
\bibfield{author}{\bibinfo{person}{Zeyuan Huang}, \bibinfo{person}{Cangjun
  Gao}, \bibinfo{person}{Yaxian Shan}, \bibinfo{person}{Haoxiang Hu},
  \bibinfo{person}{Qingkun Li}, \bibinfo{person}{Xiaoming Deng},
  \bibinfo{person}{Cuixia Ma}, \bibinfo{person}{Yu-Kun Lai},
  \bibinfo{person}{Yong-Jin Liu}, \bibinfo{person}{Feng Tian},
  \bibinfo{person}{Guozhong Dai}, {and} \bibinfo{person}{Hongan Wang}.}
  \bibinfo{year}{2025}\natexlab{}.
\newblock \showarticletitle{{SketchGPT}: A Sketch-Based Multimodal Interface
  for Application-Agnostic {LLM} Interaction}. In
  \bibinfo{booktitle}{\emph{Proceedings of the 38th Annual ACM Symposium on
  User Interface Software and Technology}}. \bibinfo{publisher}{ACM},
  \bibinfo{pages}{1--18}.
\newblock
\href{https://doi.org/10.1145/3746059.3747598}{doi:\nolinkurl{10.1145/3746059.3747598}}


\bibitem[Huynh et~al\mbox{.}(2025)]%
        {huynh2025spatial}
\bibfield{author}{\bibinfo{person}{Khanh Huynh}, \bibinfo{person}{Jeremy
  Dillmann}, {and} \bibinfo{person}{Sven Mayer}.}
  \bibinfo{year}{2025}\natexlab{}.
\newblock \showarticletitle{Spatial Referencing for Large Language Models in
  Automotive Navigation Tasks}. In \bibinfo{booktitle}{\emph{Proceedings of the
  24th International Conference on Mobile and Ubiquitous Multimedia}}.
  \bibinfo{publisher}{ACM}, \bibinfo{pages}{146--157}.
\newblock
\href{https://doi.org/10.1145/3771882.3771917}{doi:\nolinkurl{10.1145/3771882.3771917}}


\bibitem[Jia et~al\mbox{.}(2025)]%
        {jia2025voice}
\bibfield{author}{\bibinfo{person}{Donggang Jia}, \bibinfo{person}{Alexandra
  Irger}, \bibinfo{person}{Lonni Besan{\c{c}}on}, \bibinfo{person}{Ond{\v{r}}ej
  Strnad}, \bibinfo{person}{Deng Luo}, \bibinfo{person}{Johanna Bj{\"o}rklund},
  \bibinfo{person}{Alexandre Kouyoumdjian}, \bibinfo{person}{Anders Ynnerman},
  {and} \bibinfo{person}{Ivan Viola}.} \bibinfo{year}{2025}\natexlab{}.
\newblock \showarticletitle{{VOICE}: Visual Oracle for Interaction,
  Conversation, and Explanation}.
\newblock \bibinfo{journal}{\emph{IEEE Transactions on Visualization and
  Computer Graphics}} \bibinfo{volume}{31}, \bibinfo{number}{10}
  (\bibinfo{year}{2025}), \bibinfo{pages}{8828--8845}.
\newblock
\href{https://doi.org/10.1109/TVCG.2025.3579956}{doi:\nolinkurl{10.1109/TVCG.2025.3579956}}


\bibitem[Jiang et~al\mbox{.}(2023)]%
        {jiang2023graphologue}
\bibfield{author}{\bibinfo{person}{Peiling Jiang}, \bibinfo{person}{Jude
  Rayan}, \bibinfo{person}{Steven~P. Dow}, {and} \bibinfo{person}{Haijun Xia}.}
  \bibinfo{year}{2023}\natexlab{}.
\newblock \showarticletitle{Graphologue: Exploring Large Language Model
  Responses with Interactive Diagrams}. In
  \bibinfo{booktitle}{\emph{Proceedings of the 36th Annual ACM Symposium on
  User Interface Software and Technology}}. \bibinfo{publisher}{ACM},
  \bibinfo{pages}{1--20}.
\newblock
\href{https://doi.org/10.1145/3586183.3606737}{doi:\nolinkurl{10.1145/3586183.3606737}}


\bibitem[Kim et~al\mbox{.}(2026)]%
        {kim2026speaking}
\bibfield{author}{\bibinfo{person}{Yujin Kim}, \bibinfo{person}{Christine~P.
  Lee}, {and} \bibinfo{person}{Bilge Mutlu}.} \bibinfo{year}{2026}\natexlab{}.
\newblock \showarticletitle{Speaking with Screens: Design Space and Guidelines
  for Informational Robot Screens}. In \bibinfo{booktitle}{\emph{Proceedings of
  the 21st ACM/IEEE International Conference on Human-Robot Interaction}}.
  \bibinfo{publisher}{ACM}, \bibinfo{pages}{385--394}.
\newblock
\href{https://doi.org/10.1145/3757279.3785557}{doi:\nolinkurl{10.1145/3757279.3785557}}


\bibitem[Kirmayr et~al\mbox{.}(2026)]%
        {kirmayr2026intermediate}
\bibfield{author}{\bibinfo{person}{Johannes Kirmayr},
  \bibinfo{person}{Raphael~Paul Wennmacher}, \bibinfo{person}{Khanh Huynh},
  \bibinfo{person}{Lukas Stappen}, \bibinfo{person}{Elisabeth Andr{\'e}}, {and}
  \bibinfo{person}{Florian Alt}.} \bibinfo{year}{2026}\natexlab{}.
\newblock \showarticletitle{{``What Are You Doing?''}: Effects of Intermediate
  Feedback from Agentic {LLM} In-Car Assistants During Multi-Step Processing}.
  In \bibinfo{booktitle}{\emph{Proceedings of the 2026 CHI Conference on Human
  Factors in Computing Systems}}. \bibinfo{publisher}{ACM},
  \bibinfo{pages}{1--18}.
\newblock
\href{https://doi.org/10.1145/3772318.3790997}{doi:\nolinkurl{10.1145/3772318.3790997}}


\bibitem[Kretzer et~al\mbox{.}(2025)]%
        {kretzer2025closing}
\bibfield{author}{\bibinfo{person}{Felix Kretzer}, \bibinfo{person}{Kristian
  Kolthoff}, \bibinfo{person}{Christian Bartelt}, \bibinfo{person}{Simone~Paolo
  Ponzetto}, {and} \bibinfo{person}{Alexander Maedche}.}
  \bibinfo{year}{2025}\natexlab{}.
\newblock \showarticletitle{Closing the Loop between User Stories and {GUI}
  Prototypes: An {LLM}-Based Assistant for Cross-Functional Integration in
  Software Development}. In \bibinfo{booktitle}{\emph{Proceedings of the 2025
  CHI Conference on Human Factors in Computing Systems}}.
  \bibinfo{publisher}{ACM}, \bibinfo{pages}{1--19}.
\newblock
\href{https://doi.org/10.1145/3706598.3713932}{doi:\nolinkurl{10.1145/3706598.3713932}}


\bibitem[Lee et~al\mbox{.}(2026)]%
        {lee2026player}
\bibfield{author}{\bibinfo{person}{Chunggi Lee}, \bibinfo{person}{Tica Lin},
  \bibinfo{person}{Yalong Yang}, {and} \bibinfo{person}{Hanspeter Pfister}.}
  \bibinfo{year}{2026}\natexlab{}.
\newblock \bibinfo{title}{Who's That Player?: Externalizing Query
  Interpretation in Spoken {XR} Sports Interaction}.
\newblock
\showeprint[arxiv]{2608.00876}~[cs.HC]
\href{https://doi.org/10.48550/arXiv.2608.00876}{doi:\nolinkurl{10.48550/arXiv.2608.00876}}


\bibitem[Lee et~al\mbox{.}(2025c)]%
        {lee2025sensible}
\bibfield{author}{\bibinfo{person}{Geonsun Lee}, \bibinfo{person}{Min Xia},
  \bibinfo{person}{Nels Numan}, \bibinfo{person}{Xun Qian},
  \bibinfo{person}{David Li}, \bibinfo{person}{Yanhe Chen},
  \bibinfo{person}{Achin Kulshrestha}, \bibinfo{person}{Ishan Chatterjee},
  \bibinfo{person}{Yinda Zhang}, \bibinfo{person}{Dinesh Manocha},
  \bibinfo{person}{David Kim}, {and} \bibinfo{person}{Ruofei Du}.}
  \bibinfo{year}{2025}\natexlab{c}.
\newblock \showarticletitle{Sensible Agent: A Framework for Unobtrusive
  Interaction with Proactive {AR} Agents}. In
  \bibinfo{booktitle}{\emph{Proceedings of the 38th Annual ACM Symposium on
  User Interface Software and Technology}}. \bibinfo{publisher}{ACM},
  \bibinfo{pages}{1--22}.
\newblock
\href{https://doi.org/10.1145/3746059.3747748}{doi:\nolinkurl{10.1145/3746059.3747748}}


\bibitem[Lee et~al\mbox{.}(2025a)]%
        {lee2025imaginatear}
\bibfield{author}{\bibinfo{person}{Jaewook Lee}, \bibinfo{person}{Filippo
  Aleotti}, \bibinfo{person}{Diego Mazala}, \bibinfo{person}{Guillermo
  Garcia-Hernando}, \bibinfo{person}{Sara Vicente},
  \bibinfo{person}{Oliver~James Johnston}, \bibinfo{person}{Isabel
  Kraus-Liang}, \bibinfo{person}{Jakub Powierza}, \bibinfo{person}{Donghoon
  Shin}, \bibinfo{person}{Jon~E. Froehlich}, \bibinfo{person}{Gabriel Brostow},
  {and} \bibinfo{person}{Jessica Van~Brummelen}.}
  \bibinfo{year}{2025}\natexlab{a}.
\newblock \showarticletitle{{ImaginateAR}: {AI}-Assisted In-Situ Authoring in
  Augmented Reality}. In \bibinfo{booktitle}{\emph{Proceedings of the 38th
  Annual ACM Symposium on User Interface Software and Technology}}.
  \bibinfo{publisher}{ACM}, \bibinfo{pages}{1--21}.
\newblock
\href{https://doi.org/10.1145/3746059.3747635}{doi:\nolinkurl{10.1145/3746059.3747635}}


\bibitem[Lee et~al\mbox{.}(2025b)]%
        {lee2025sketchpad}
\bibfield{author}{\bibinfo{person}{Jimin Lee}, \bibinfo{person}{Steven-Shine
  Chen}, {and} \bibinfo{person}{Paul~Pu Liang}.}
  \bibinfo{year}{2025}\natexlab{b}.
\newblock \showarticletitle{Interactive Sketchpad: A Multimodal Tutoring System
  for Collaborative, Visual Problem-Solving}. In
  \bibinfo{booktitle}{\emph{Extended Abstracts of the 2025 CHI Conference on
  Human Factors in Computing Systems}}. \bibinfo{publisher}{ACM},
  \bibinfo{pages}{1--14}.
\newblock
\href{https://doi.org/10.1145/3706599.3719790}{doi:\nolinkurl{10.1145/3706599.3719790}}


\bibitem[Lee et~al\mbox{.}(2024)]%
        {lee2024gazepointar}
\bibfield{author}{\bibinfo{person}{Jaewook Lee}, \bibinfo{person}{Jun Wang},
  \bibinfo{person}{Elizabeth Brown}, \bibinfo{person}{Liam Chu},
  \bibinfo{person}{Sebastian~S. Rodriguez}, {and} \bibinfo{person}{Jon~E.
  Froehlich}.} \bibinfo{year}{2024}\natexlab{}.
\newblock \showarticletitle{{GazePointAR}: A Context-Aware Multimodal Voice
  Assistant for Pronoun Disambiguation in Wearable Augmented Reality}. In
  \bibinfo{booktitle}{\emph{Proceedings of the 2024 CHI Conference on Human
  Factors in Computing Systems}}. \bibinfo{publisher}{ACM},
  \bibinfo{pages}{1--20}.
\newblock
\href{https://doi.org/10.1145/3613904.3642230}{doi:\nolinkurl{10.1145/3613904.3642230}}


\bibitem[Li and Jiang(2026)]%
        {li2026comap}
\bibfield{author}{\bibinfo{person}{Ruijia Li} {and} \bibinfo{person}{Bo
  Jiang}.} \bibinfo{year}{2026}\natexlab{}.
\newblock \showarticletitle{Thinking in Graphs with {CoMAP}: A Shared Visual
  Workspace for Designing Project-Based Learning}. In
  \bibinfo{booktitle}{\emph{Proceedings of the 2026 CHI Conference on Human
  Factors in Computing Systems}}. \bibinfo{publisher}{ACM},
  \bibinfo{pages}{1--25}.
\newblock
\href{https://doi.org/10.1145/3772318.3790273}{doi:\nolinkurl{10.1145/3772318.3790273}}


\bibitem[Li et~al\mbox{.}(2026)]%
        {li2026orality}
\bibfield{author}{\bibinfo{person}{Wengxi Li}, \bibinfo{person}{Jingze Tian},
  {and} \bibinfo{person}{Can Liu}.} \bibinfo{year}{2026}\natexlab{}.
\newblock \showarticletitle{Orality: A Semantic Canvas for Externalizing and
  Clarifying Thoughts with Speech}. In \bibinfo{booktitle}{\emph{Proceedings of
  the 2026 CHI Conference on Human Factors in Computing Systems}}.
  \bibinfo{publisher}{ACM}, \bibinfo{pages}{1--29}.
\newblock
\href{https://doi.org/10.1145/3772318.3791713}{doi:\nolinkurl{10.1145/3772318.3791713}}


\bibitem[Liu et~al\mbox{.}(2025)]%
        {liu2025bargein}
\bibfield{author}{\bibinfo{person}{Chao Liu}, \bibinfo{person}{Mingyang Su},
  \bibinfo{person}{Yan Xiang}, \bibinfo{person}{Yuru Huang},
  \bibinfo{person}{Yiqian Yang}, \bibinfo{person}{Kang Zhang}, {and}
  \bibinfo{person}{Mingming Fan}.} \bibinfo{year}{2025}\natexlab{}.
\newblock \showarticletitle{Toward Enabling Natural Conversation with Older
  Adults via the Design of {LLM}-Powered Voice Agents that Support
  Interruptions and Backchannels}. In \bibinfo{booktitle}{\emph{Proceedings of
  the 2025 CHI Conference on Human Factors in Computing Systems}}.
  \bibinfo{publisher}{ACM}, \bibinfo{pages}{1--22}.
\newblock
\href{https://doi.org/10.1145/3706598.3714228}{doi:\nolinkurl{10.1145/3706598.3714228}}


\bibitem[Liu et~al\mbox{.}(2023)]%
        {liu2023visualcaptions}
\bibfield{author}{\bibinfo{person}{Xingyu~Bruce Liu}, \bibinfo{person}{Vladimir
  Kirilyuk}, \bibinfo{person}{Xiuxiu Yuan}, \bibinfo{person}{Alex Olwal},
  \bibinfo{person}{Peggy Chi}, \bibinfo{person}{Xiang~Anthony Chen}, {and}
  \bibinfo{person}{Ruofei Du}.} \bibinfo{year}{2023}\natexlab{}.
\newblock \showarticletitle{Visual Captions: Augmenting Verbal Communication
  with On-the-Fly Visuals}. In \bibinfo{booktitle}{\emph{Proceedings of the
  2023 CHI Conference on Human Factors in Computing Systems}}.
  \bibinfo{publisher}{ACM}, \bibinfo{pages}{1--20}.
\newblock
\href{https://doi.org/10.1145/3544548.3581566}{doi:\nolinkurl{10.1145/3544548.3581566}}


\bibitem[Liu et~al\mbox{.}(2026)]%
        {liu2026amiwrite}
\bibfield{author}{\bibinfo{person}{Ziyi Liu}, \bibinfo{person}{Yuzhao Chen},
  \bibinfo{person}{Haoyu Ji}, \bibinfo{person}{Runlin Duan},
  \bibinfo{person}{Zhengzhe Zhu}, \bibinfo{person}{Xiyun Hu},
  \bibinfo{person}{Kylie Peppler}, {and} \bibinfo{person}{Karthik Ramani}.}
  \bibinfo{year}{2026}\natexlab{}.
\newblock \showarticletitle{{AmIWrite}: Exploring Scalable One-on-One
  Handwriting-Based Tutoring for Mathematical Problem-Solving with an
  {LLM}-Powered {AI} Tutor}. In \bibinfo{booktitle}{\emph{Proceedings of the
  2026 CHI Conference on Human Factors in Computing Systems}}.
  \bibinfo{publisher}{ACM}, \bibinfo{pages}{1--28}.
\newblock
\href{https://doi.org/10.1145/3772318.3790935}{doi:\nolinkurl{10.1145/3772318.3790935}}


\bibitem[Ma and Cao(2019)]%
        {ma2019design}
\bibfield{author}{\bibinfo{person}{Ke Ma} {and} \bibinfo{person}{Jing Cao}.}
  \bibinfo{year}{2019}\natexlab{}.
\newblock \showarticletitle{Design Pattern as a Practical Tool for Designing
  Adaptive Interactions Connecting Human and Social Robots}. In
  \bibinfo{booktitle}{\emph{Intelligent Human Systems Integration 2019}}
  \emph{(\bibinfo{series}{Advances in Intelligent Systems and Computing},
  Vol.~\bibinfo{volume}{903})}, \bibfield{editor}{\bibinfo{person}{Waldemar
  Karwowski} {and} \bibinfo{person}{Tareq Ahram}} (Eds.).
  \bibinfo{publisher}{Springer International Publishing},
  \bibinfo{address}{Cham}, \bibinfo{pages}{613--617}.
\newblock
\href{https://doi.org/10.1007/978-3-030-11051-2_93}{doi:\nolinkurl{10.1007/978-3-030-11051-2_93}}


\bibitem[Mahmood et~al\mbox{.}(2025)]%
        {mahmood2025voice}
\bibfield{author}{\bibinfo{person}{Amama Mahmood}, \bibinfo{person}{Junxiang
  Wang}, \bibinfo{person}{Bingsheng Yao}, \bibinfo{person}{Dakuo Wang}, {and}
  \bibinfo{person}{Chien-Ming Huang}.} \bibinfo{year}{2025}\natexlab{}.
\newblock \showarticletitle{User Interaction Patterns and Breakdowns in
  Conversing with {LLM}-Powered Voice Assistants}.
\newblock \bibinfo{journal}{\emph{International Journal of Human-Computer
  Studies}}  \bibinfo{volume}{195} (\bibinfo{year}{2025}),
  \bibinfo{pages}{103406}.
\newblock
\href{https://doi.org/10.1016/j.ijhcs.2024.103406}{doi:\nolinkurl{10.1016/j.ijhcs.2024.103406}}


\bibitem[Maslych et~al\mbox{.}(2025)]%
        {maslych2025mitigating}
\bibfield{author}{\bibinfo{person}{Mykola Maslych},
  \bibinfo{person}{Mohammadreza Katebi}, \bibinfo{person}{Christopher Lee},
  \bibinfo{person}{Yahya Hmaiti}, \bibinfo{person}{Amirpouya Ghasemaghaei},
  \bibinfo{person}{Christian Pumarada}, \bibinfo{person}{Janneese Palmer},
  \bibinfo{person}{Esteban Segarra~Martinez}, \bibinfo{person}{Marco Emporio},
  \bibinfo{person}{Warren Snipes}, \bibinfo{person}{Ryan~P. McMahan}, {and}
  \bibinfo{person}{Joseph~J. LaViola, Jr.}} \bibinfo{year}{2025}\natexlab{}.
\newblock \showarticletitle{Mitigating Response Delays in Free-Form
  Conversations with {LLM}-powered Intelligent Virtual Agents}. In
  \bibinfo{booktitle}{\emph{Proceedings of the 7th ACM Conference on
  Conversational User Interfaces}}. \bibinfo{publisher}{ACM},
  \bibinfo{pages}{1--15}.
\newblock
\href{https://doi.org/10.1145/3719160.3736636}{doi:\nolinkurl{10.1145/3719160.3736636}}


\bibitem[Mayer and Moreno(2003)]%
        {mayer2003nine}
\bibfield{author}{\bibinfo{person}{Richard~E. Mayer} {and}
  \bibinfo{person}{Roxana Moreno}.} \bibinfo{year}{2003}\natexlab{}.
\newblock \showarticletitle{Nine Ways to Reduce Cognitive Load in Multimedia
  Learning}.
\newblock \bibinfo{journal}{\emph{Educational Psychologist}}
  \bibinfo{volume}{38}, \bibinfo{number}{1} (\bibinfo{year}{2003}),
  \bibinfo{pages}{43--52}.
\newblock
\href{https://doi.org/10.1207/S15326985EP3801_6}{doi:\nolinkurl{10.1207/S15326985EP3801_6}}


\bibitem[{Microsoft}(2026)]%
        {microsoft2026huabu}
\bibfield{author}{\bibinfo{person}{{Microsoft}}.}
  \bibinfo{year}{2026}\natexlab{}.
\newblock \bibinfo{title}{{Huabu}: Where You and Your Agents Think Together}.
\newblock
\urldef\tempurl%
\url{https://github.com/microsoft/Huabu}
\showURL{%
\tempurl}


\bibitem[Oh et~al\mbox{.}(2024)]%
        {oh2024voicon}
\bibfield{author}{\bibinfo{person}{Jeesun Oh}, \bibinfo{person}{Nayeong Kim},
  \bibinfo{person}{Yuchun Yan}, {and} \bibinfo{person}{Sangsu Lee}.}
  \bibinfo{year}{2024}\natexlab{}.
\newblock \showarticletitle{{VOICON}: Geometric Motion-Based Visual Feedback in
  Voice User Interface}. In \bibinfo{booktitle}{\emph{Proceedings of the 2024
  ACM Designing Interactive Systems Conference}}. \bibinfo{publisher}{ACM},
  \bibinfo{pages}{102--115}.
\newblock
\href{https://doi.org/10.1145/3643834.3660741}{doi:\nolinkurl{10.1145/3643834.3660741}}


\bibitem[Porcheron et~al\mbox{.}(2018)]%
        {porcheron2018voice}
\bibfield{author}{\bibinfo{person}{Martin Porcheron}, \bibinfo{person}{Joel~E.
  Fischer}, \bibinfo{person}{Stuart Reeves}, {and} \bibinfo{person}{Sarah
  Sharples}.} \bibinfo{year}{2018}\natexlab{}.
\newblock \showarticletitle{Voice Interfaces in Everyday Life}. In
  \bibinfo{booktitle}{\emph{Proceedings of the 2018 CHI Conference on Human
  Factors in Computing Systems}}. \bibinfo{publisher}{ACM},
  \bibinfo{pages}{1--12}.
\newblock
\href{https://doi.org/10.1145/3173574.3174214}{doi:\nolinkurl{10.1145/3173574.3174214}}


\bibitem[Prasad and Mahapatra(2026)]%
        {prasad2026whiteboard}
\bibfield{author}{\bibinfo{person}{Suraj Prasad} {and} \bibinfo{person}{Pinak
  Mahapatra}.} \bibinfo{year}{2026}\natexlab{}.
\newblock \bibinfo{title}{Speech-Synchronized Whiteboard Generation via
  {VLM}-Driven Structured Drawing Representations}.
\newblock
\showeprint[arxiv]{2603.25870}~[cs.CV]
\href{https://doi.org/10.48550/arXiv.2603.25870}{doi:\nolinkurl{10.48550/arXiv.2603.25870}}


\bibitem[Reicherts et~al\mbox{.}(2022)]%
        {reicherts2022good}
\bibfield{author}{\bibinfo{person}{Leon Reicherts}, \bibinfo{person}{Yvonne
  Rogers}, \bibinfo{person}{Licia Capra}, \bibinfo{person}{Ethan Wood},
  \bibinfo{person}{Tu~Dinh Duong}, {and} \bibinfo{person}{Neil~J. Sebire}.}
  \bibinfo{year}{2022}\natexlab{}.
\newblock \showarticletitle{It's Good to Talk: A Comparison of Using Voice
  Versus Screen-Based Interactions for Agent-Assisted Tasks}.
\newblock \bibinfo{journal}{\emph{ACM Transactions on Computer-Human
  Interaction}} \bibinfo{volume}{29}, \bibinfo{number}{3}, Article
  \bibinfo{articleno}{25} (\bibinfo{year}{2022}), \bibinfo{numpages}{41}~pages.
\newblock
\href{https://doi.org/10.1145/3484221}{doi:\nolinkurl{10.1145/3484221}}


\bibitem[Rosenberg et~al\mbox{.}(2024)]%
        {rosenberg2024drawtalking}
\bibfield{author}{\bibinfo{person}{Karl~Toby Rosenberg},
  \bibinfo{person}{Rubaiat~Habib Kazi}, \bibinfo{person}{Li-Yi Wei},
  \bibinfo{person}{Haijun Xia}, {and} \bibinfo{person}{Ken Perlin}.}
  \bibinfo{year}{2024}\natexlab{}.
\newblock \showarticletitle{{DrawTalking}: Building Interactive Worlds by
  Sketching and Speaking}. In \bibinfo{booktitle}{\emph{Proceedings of the 37th
  Annual ACM Symposium on User Interface Software and Technology}}.
  \bibinfo{publisher}{ACM}, \bibinfo{pages}{1--25}.
\newblock
\href{https://doi.org/10.1145/3654777.3676334}{doi:\nolinkurl{10.1145/3654777.3676334}}


\bibitem[Scaife and Rogers(1996)]%
        {scaife1996external}
\bibfield{author}{\bibinfo{person}{Mike Scaife} {and} \bibinfo{person}{Yvonne
  Rogers}.} \bibinfo{year}{1996}\natexlab{}.
\newblock \showarticletitle{External Cognition: How Do Graphical
  Representations Work?}
\newblock \bibinfo{journal}{\emph{International Journal of Human-Computer
  Studies}} \bibinfo{volume}{45}, \bibinfo{number}{2} (\bibinfo{year}{1996}),
  \bibinfo{pages}{185--213}.
\newblock
\href{https://doi.org/10.1006/ijhc.1996.0048}{doi:\nolinkurl{10.1006/ijhc.1996.0048}}


\bibitem[Shen et~al\mbox{.}(2024)]%
        {shen2024dataplayer}
\bibfield{author}{\bibinfo{person}{Leixian Shen}, \bibinfo{person}{Yizhi
  Zhang}, \bibinfo{person}{Haidong Zhang}, {and} \bibinfo{person}{Yun Wang}.}
  \bibinfo{year}{2024}\natexlab{}.
\newblock \showarticletitle{Data Player: Automatic Generation of Data Videos
  with Narration-Animation Interplay}.
\newblock \bibinfo{journal}{\emph{IEEE Transactions on Visualization and
  Computer Graphics}} \bibinfo{volume}{30}, \bibinfo{number}{1}
  (\bibinfo{year}{2024}), \bibinfo{pages}{109--119}.
\newblock
\href{https://doi.org/10.1109/TVCG.2023.3327197}{doi:\nolinkurl{10.1109/TVCG.2023.3327197}}


\bibitem[Shi et~al\mbox{.}(2026)]%
        {shi2026drawing}
\bibfield{author}{\bibinfo{person}{Weiyan Shi}, \bibinfo{person}{Dorien
  Herremans}, {and} \bibinfo{person}{Kenny Tsu~Wei Choo}.}
  \bibinfo{year}{2026}\natexlab{}.
\newblock \bibinfo{title}{When Drawing Is Not Enough: Exploring Spontaneous
  Speech with Sketch for Intent Alignment in Multimodal {LLM}s}.
\newblock
\showeprint[arxiv]{2604.11964}~[cs.HC]
\href{https://doi.org/10.48550/arXiv.2604.11964}{doi:\nolinkurl{10.48550/arXiv.2604.11964}}


\bibitem[Sorokin et~al\mbox{.}(2025)]%
        {sorokin2025voicegui}
\bibfield{author}{\bibinfo{person}{Lenja Sorokin}, \bibinfo{person}{Khanh
  Huynh}, \bibinfo{person}{Malin Eiband}, \bibinfo{person}{Lukas Stappen},
  {and} \bibinfo{person}{Jeremy Dillmann}.} \bibinfo{year}{2025}\natexlab{}.
\newblock \showarticletitle{Collaborating with {LLM}s Through a Voice and
  Graphical User Interface}. In \bibinfo{booktitle}{\emph{Adjunct Proceedings
  of the 27th International Conference on Mobile Human-Computer Interaction}}.
  \bibinfo{publisher}{ACM}, \bibinfo{pages}{1--8}.
\newblock
\href{https://doi.org/10.1145/3737821.3749555}{doi:\nolinkurl{10.1145/3737821.3749555}}


\bibitem[Suh et~al\mbox{.}(2024)]%
        {suh2024luminate}
\bibfield{author}{\bibinfo{person}{Sangho Suh}, \bibinfo{person}{Meng Chen},
  \bibinfo{person}{Bryan Min}, \bibinfo{person}{Toby Jia-Jun Li}, {and}
  \bibinfo{person}{Haijun Xia}.} \bibinfo{year}{2024}\natexlab{}.
\newblock \showarticletitle{Luminate: Structured Generation and Exploration of
  Design Space with Large Language Models for Human--AI Co-Creation}. In
  \bibinfo{booktitle}{\emph{Proceedings of the 2024 CHI Conference on Human
  Factors in Computing Systems}}. \bibinfo{publisher}{ACM},
  \bibinfo{pages}{1--26}.
\newblock
\href{https://doi.org/10.1145/3613904.3642400}{doi:\nolinkurl{10.1145/3613904.3642400}}


\bibitem[Suh et~al\mbox{.}(2023)]%
        {suh2023sensecape}
\bibfield{author}{\bibinfo{person}{Sangho Suh}, \bibinfo{person}{Bryan Min},
  \bibinfo{person}{Srishti Palani}, {and} \bibinfo{person}{Haijun Xia}.}
  \bibinfo{year}{2023}\natexlab{}.
\newblock \showarticletitle{Sensecape: Enabling Multilevel Exploration and
  Sensemaking with Large Language Models}. In
  \bibinfo{booktitle}{\emph{Proceedings of the 36th Annual ACM Symposium on
  User Interface Software and Technology}}. \bibinfo{publisher}{ACM},
  \bibinfo{pages}{1--18}.
\newblock
\href{https://doi.org/10.1145/3586183.3606756}{doi:\nolinkurl{10.1145/3586183.3606756}}


\bibitem[Tan(2021)]%
        {tan2021bansho}
\bibfield{author}{\bibinfo{person}{Shirley Tan}.}
  \bibinfo{year}{2021}\natexlab{}.
\newblock \showarticletitle{{Bansho} as Part of Lesson and Lesson Study: From
  the Origins to the Present}.
\newblock \bibinfo{journal}{\emph{International Journal for Lesson \& Learning
  Studies}} \bibinfo{volume}{10}, \bibinfo{number}{4} (\bibinfo{year}{2021}),
  \bibinfo{pages}{378--392}.
\newblock
\href{https://doi.org/10.1108/IJLLS-09-2021-0076}{doi:\nolinkurl{10.1108/IJLLS-09-2021-0076}}


\bibitem[Tan et~al\mbox{.}(2023a)]%
        {tan2023representations}
\bibfield{author}{\bibinfo{person}{Shirley Tan}, \bibinfo{person}{St{\'e}phane
  Clivaz}, {and} \bibinfo{person}{Masanobu Sakamoto}.}
  \bibinfo{year}{2023}\natexlab{a}.
\newblock \showarticletitle{Presenting Multiple Representations at the
  Chalkboard: {Bansho} Analysis of a Japanese Mathematics Classroom}.
\newblock \bibinfo{journal}{\emph{Journal of Education for Teaching}}
  \bibinfo{volume}{49}, \bibinfo{number}{4} (\bibinfo{year}{2023}),
  \bibinfo{pages}{630--647}.
\newblock
\href{https://doi.org/10.1080/02607476.2022.2150538}{doi:\nolinkurl{10.1080/02607476.2022.2150538}}


\bibitem[Tan et~al\mbox{.}(2023b)]%
        {tan2023decision}
\bibfield{author}{\bibinfo{person}{Shirley Tan}, \bibinfo{person}{Shiho
  Nozaki}, \bibinfo{person}{Hongxue Fu}, {and} \bibinfo{person}{Yoshiaki
  Shibata}.} \bibinfo{year}{2023}\natexlab{b}.
\newblock \showarticletitle{The Principles of Teacher's Decision-Making in
  Japanese Board Writing ({Bansho}) Process}.
\newblock \bibinfo{journal}{\emph{Asia Pacific Journal of Education}}
  \bibinfo{volume}{43}, \bibinfo{number}{1} (\bibinfo{year}{2023}),
  \bibinfo{pages}{236--251}.
\newblock
\href{https://doi.org/10.1080/02188791.2021.1924119}{doi:\nolinkurl{10.1080/02188791.2021.1924119}}


\bibitem[{Thesys}(2026)]%
        {livekit2026genui}
\bibfield{author}{\bibinfo{person}{{Thesys}}.} \bibinfo{year}{2026}\natexlab{}.
\newblock \bibinfo{title}{LiveKit Generative {UI} Voice Agent}.
\newblock
\urldef\tempurl%
\url{https://github.com/thesysdev/voice-agent-generativeui}
\showURL{%
\tempurl}


\bibitem[{THU-MAIC}(2026)]%
        {openmaic2026}
\bibfield{author}{\bibinfo{person}{{THU-MAIC}}.}
  \bibinfo{year}{2026}\natexlab{}.
\newblock \bibinfo{title}{{OpenMAIC}: Open Multi-Agent Interactive Classroom}.
\newblock
\urldef\tempurl%
\url{https://github.com/THU-MAIC/OpenMAIC}
\showURL{%
\tempurl}


\bibitem[Vaithilingam et~al\mbox{.}(2024)]%
        {vaithilingam2024dynavis}
\bibfield{author}{\bibinfo{person}{Priyan Vaithilingam},
  \bibinfo{person}{Elena~L. Glassman}, \bibinfo{person}{Jeevana~Priya Inala},
  {and} \bibinfo{person}{Chenglong Wang}.} \bibinfo{year}{2024}\natexlab{}.
\newblock \showarticletitle{{DynaVis}: Dynamically Synthesized {UI} Widgets for
  Visualization Editing}. In \bibinfo{booktitle}{\emph{Proceedings of the 2024
  CHI Conference on Human Factors in Computing Systems}}.
  \bibinfo{publisher}{ACM}, \bibinfo{pages}{1--17}.
\newblock
\href{https://doi.org/10.1145/3613904.3642639}{doi:\nolinkurl{10.1145/3613904.3642639}}


\bibitem[van Dam(2025)]%
        {vandam2025architecture}
\bibfield{author}{\bibinfo{person}{Hans G.~W. van Dam}.}
  \bibinfo{year}{2025}\natexlab{}.
\newblock \bibinfo{title}{A Multimodal {GUI} Architecture for Interfacing with
  {LLM}-Based Conversational Assistants}.
\newblock
\showeprint[arxiv]{2510.06223}~[cs.HC]
\href{https://doi.org/10.48550/arXiv.2510.06223}{doi:\nolinkurl{10.48550/arXiv.2510.06223}}


\bibitem[Veluri et~al\mbox{.}(2024)]%
        {veluri2024synchronous}
\bibfield{author}{\bibinfo{person}{Bandhav Veluri},
  \bibinfo{person}{Benjamin~N. Peloquin}, \bibinfo{person}{Bokai Yu},
  \bibinfo{person}{Hongyu Gong}, {and} \bibinfo{person}{Shyamnath Gollakota}.}
  \bibinfo{year}{2024}\natexlab{}.
\newblock \showarticletitle{Beyond Turn-Based Interfaces: Synchronous {LLM}s as
  Full-Duplex Dialogue Agents}. In \bibinfo{booktitle}{\emph{Proceedings of the
  2024 Conference on Empirical Methods in Natural Language Processing}}.
  \bibinfo{publisher}{Association for Computational Linguistics},
  \bibinfo{pages}{21390--21402}.
\newblock
\href{https://doi.org/10.18653/v1/2024.emnlp-main.1192}{doi:\nolinkurl{10.18653/v1/2024.emnlp-main.1192}}


\bibitem[Vu et~al\mbox{.}(2024)]%
        {vu2024gptvoicetasker}
\bibfield{author}{\bibinfo{person}{Minh~Duc Vu}, \bibinfo{person}{Han Wang},
  \bibinfo{person}{Jieshan Chen}, \bibinfo{person}{Zhuang Li},
  \bibinfo{person}{Shengdong Zhao}, \bibinfo{person}{Zhenchang Xing}, {and}
  \bibinfo{person}{Chunyang Chen}.} \bibinfo{year}{2024}\natexlab{}.
\newblock \showarticletitle{{GPTVoiceTasker}: Advancing Multi-Step Mobile Task
  Efficiency Through Dynamic Interface Exploration and Learning}. In
  \bibinfo{booktitle}{\emph{Proceedings of the 37th Annual ACM Symposium on
  User Interface Software and Technology}}. \bibinfo{publisher}{ACM},
  \bibinfo{pages}{1--17}.
\newblock
\href{https://doi.org/10.1145/3654777.3676356}{doi:\nolinkurl{10.1145/3654777.3676356}}


\bibitem[Zhong et~al\mbox{.}(2025)]%
        {zhong2025textilebot}
\bibfield{author}{\bibinfo{person}{Shu Zhong}, \bibinfo{person}{Elia Gatti},
  \bibinfo{person}{James Hardwick}, \bibinfo{person}{Miriam Ribul},
  \bibinfo{person}{Youngjun Cho}, {and} \bibinfo{person}{Marianna Obrist}.}
  \bibinfo{year}{2025}\natexlab{}.
\newblock \showarticletitle{{LLM}-Mediated Domain-Specific Voice Agents: The
  Case of {TextileBot}}.
\newblock \bibinfo{journal}{\emph{Behaviour \& Information Technology}}
  (\bibinfo{year}{2025}), \bibinfo{pages}{1--33}.
\newblock
\href{https://doi.org/10.1080/0144929X.2025.2456667}{doi:\nolinkurl{10.1080/0144929X.2025.2456667}}


\end{thebibliography}

\appendix

\section{\texorpdfstring{Example Speech--Canvas Action Planner Output}{Example Speech--Canvas Action Planner Output}}
\label{app:planner-output-example}

The following illustrative example shows one possible abbreviated Planner output.
Consider a user who asks, ``How might social feedback affect repeated habit
attempts, and where could I focus a study?'' The Canvas is initially empty, so
the example creates a Visual Representation without links to earlier objects.

Speech explains the possible relationship and proposes a direction for inquiry.
The Representation makes the sequence available for inspection, while a Comment
preserves the proposed intervention point on the feedback element. Temporary
Attention Guidance connects the spoken reference to that element. The next Beat
reuses the same element as Speech develops the study question, illustrating how
visible content can support subsequent speech without being recreated.

The \texttt{preface} field contains Speech only. The first Beat creates the
Representation, then attaches Attention Guidance and Annotation to the same
Speech span. Later Beats are omitted. The example illustrates content coordination
and serialization; it is not an observed study episode or a required response template.

\begin{lstlisting}[style=responseplan,float=!htbp]
{
  preface:
    "Let's map that relationship first, then narrow the study question.",
  beats: [
    { spans: [
      { canvas: [{
          type: "representation.create",
          id: "habit-feedback-loop",
          title: "Habit and Social Feedback Loop",
          sources: [],
          source: "flowchart LR\nattempt[\"Habit attempt\"] --> feedback[\"Social feedback\"]\nfeedback --> next[\"Next attempt\"]"
        }],
        speech:
          "A habit attempt can elicit social feedback, which may shape the next attempt." },
      { canvas: [
        {
          type: "attention.show",
          targets: [{
            objectId: "habit-feedback-loop",
            targetId: "feedback"
          }]
        },
        {
          type: "annotation.add",
          id: "potential-intervention",
          target: {
            objectId: "habit-feedback-loop",
            targetId: "feedback"
          },
          comment: "Potential intervention point"
        }
      ],
        speech:
          "The feedback step is a promising intervention point." }
    ] },
    { spans: [
      { canvas: [{
          type: "attention.show",
          targets: [{
            objectId: "habit-feedback-loop",
            targetId: "feedback"
          }]
        }],
        speech:
          "The key question is what kind of feedback changes the next attempt." },
      { speech:
          "That gives us a focused direction for the study." }
    ] },
    ...
  ]
}
\end{lstlisting}

\section{Example Canvas Snapshots from Deployment Sessions}
\label{app:session-examples}

These three examples come from Sessions discussed in
Section~\ref{sec:findings-rq2}. They retain the original image text and use
English captions based on the interaction logs.

\begin{figure*}[t]
  \centering
  \includegraphics[width=\textwidth]{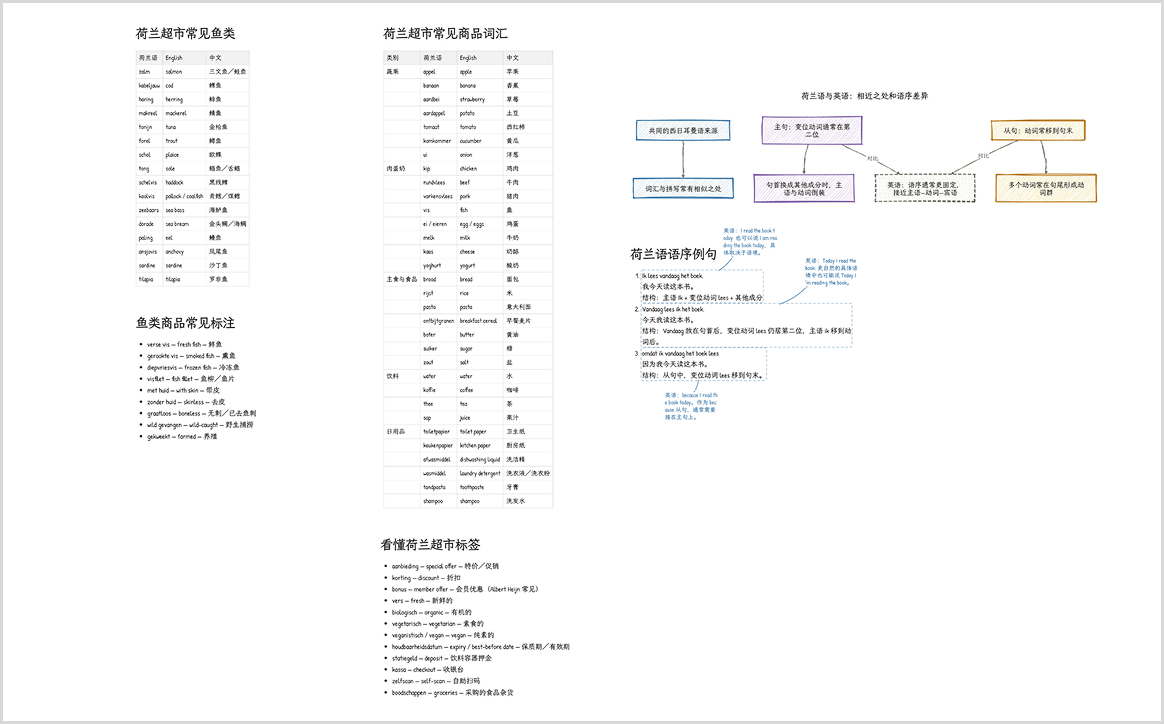}
  \caption{P04 requested written Dutch examples and English equivalents;
  the agent created a Note and added Annotations.}
  \Description{Vocabulary tables accompany a Dutch word-order diagram and
  a Note with three example sentences and Chinese meanings. Attached
  annotations provide English equivalents and word-order explanations.}
  \label{fig:session-dutch}
\end{figure*}

\begin{figure*}[t]
  \centering
  \includegraphics[width=\textwidth]{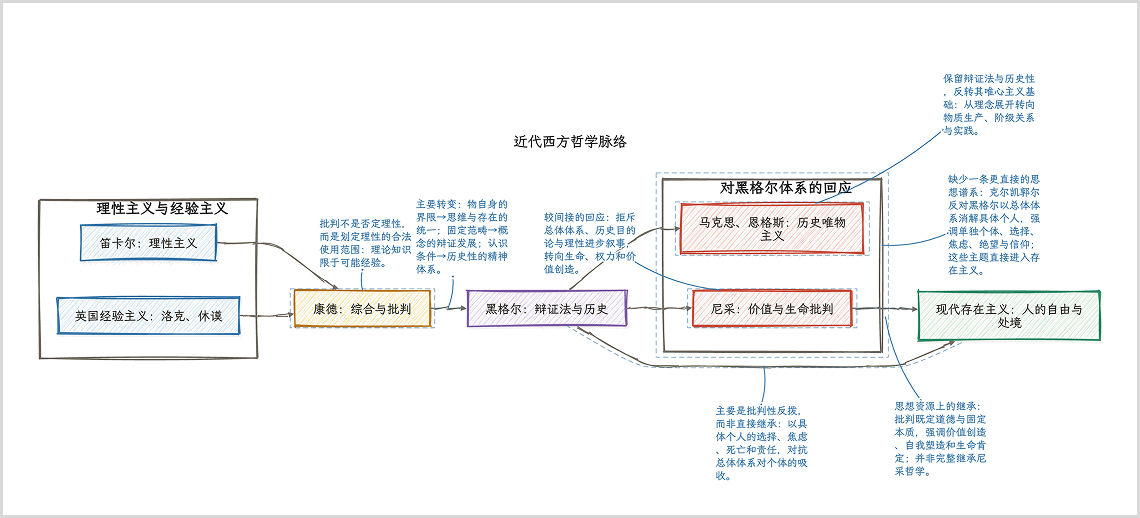}
  \caption{P16 questioned arrows toward existentialism; the agent explained
  the relationships through speech and annotated the existing diagram.}
  \Description{A diagram links rationalism and empiricism, Kant, Hegel, Marx,
  Nietzsche, and existentialism. Annotations distinguish critical responses
  and intellectual influences from direct inheritance.}
  \label{fig:session-philosophy}
\end{figure*}

\begin{figure}[htbp]
  \centering
  \includegraphics[width=\linewidth]{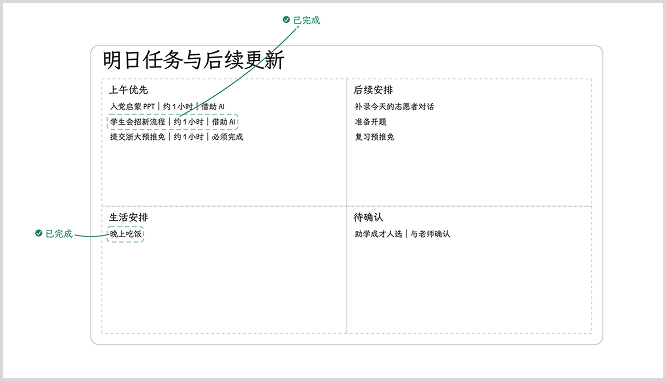}
  \caption{P05 returned after approximately 35 hours and reported two
  completed activities; the agent added completion Badges to the existing
  Discussion Frame.}
  \Description{A four-area task Frame organizes priorities, follow-up plans,
  everyday activities, and unconfirmed items. Completed Badges point to a
  recruitment poster task and dinner; other entries remain unmarked.}
  \label{fig:session-task-updates}
\end{figure}

\section{Questionnaire Instruments}
\label{app:questionnaires}

\subsection{Session Reflection}
\label{app:session-reflection}

After a substantive deployment Session, participants were invited to provide
an overall experience rating and one combined response to three open-ended
prompts. The English wording used in the interface was:
\begin{enumerate}
  \item \emph{Overall experience rating:} ``Considering only this discussion,
  how was your overall experience?'' The five response options were
  1 (Poor), 2 (Fair), 3 (Okay), 4 (Good), and 5 (Great).
  \item \emph{Fixed opening prompt:} ``Briefly, did this session help with the
  topic you wanted to discuss? What was most helpful or most difficult?''
  \item \emph{Rotating prompt:} One of the following four prompts was selected
  for each Session and remained the same whenever its feedback panel was opened:
  \begin{enumerate}
    \item ``Did anything shown on the canvas help you understand or move the
    topic forward? Was any content unnecessary or distracting?''
    \item ``As the discussion continued, the system gradually added and
    organized content on the canvas. Did that process feel coherent and build
    on the earlier discussion? Did anything feel disconnected, repetitive,
    or unclear?''
    \item ``As the system talked with you while presenting or adjusting
    content on the canvas, did the two feel naturally coordinated? Was there
    a moment when the experience felt abrupt, hard to follow, or out of sync?''
    \item ``Was there a moment when you wanted to control the canvas directly
    instead of asking the system through conversation---for example, to edit,
    move, delete, or add something yourself? If so, what did you want to do?''
  \end{enumerate}
  \item \emph{Fixed closing prompt:} ``Besides that, is there anything else
  you would like to share about this session? Any thoughts about the discussion,
  voice, canvas, or overall experience are welcome.''
\end{enumerate}

\subsection{Post-Deployment Questionnaire}

The post-deployment questionnaire contained three blocks. Agreement items used
a five-point scale from \emph{strongly disagree} to \emph{strongly agree}.

\subsubsection{System Usability Scale}

We administered the standard SUS items with their original five-point response
scale \cite{brooke1996sus}:
\begin{enumerate}
  \item I think that I would like to use this system frequently.
  \item I found the system unnecessarily complex.
  \item I thought the system was easy to use.
  \item I think that I would need the support of a technical person to be able
  to use this system.
  \item I found the various functions in this system were well integrated.
  \item I thought there was too much inconsistency in this system.
  \item I would imagine that most people would learn to use this system very
  quickly.
  \item I found the system very cumbersome to use.
  \item I felt very confident using the system.
  \item I needed to learn a lot of things before I could get going with this
  system.
\end{enumerate}

\subsubsection{Speech--Canvas Interaction Experience}

Participants were instructed to consider their overall experience across the
five-day deployment:
\begin{enumerate}
  \setcounter{enumi}{10}
  \item I could understand the relationship between what \textsc{VoCa} said
  and the changes on the Canvas.
  \item \textsc{VoCa}'s speech and Canvas content complemented each other
  rather than repeating unnecessarily or conflicting.
  \item The timing of Canvas changes made it easy to relate them to what I was
  hearing.
  \item In Sessions with more content, the Canvas remained clear and easy to
  use.
  \item When I tried to guide or correct \textsc{VoCa}, the system usually
  adjusted its subsequent response as I intended.
  \item At times, I found it difficult to keep up with \textsc{VoCa}'s speech
  and Canvas changes at the same time.
\end{enumerate}

\subsubsection{Voice-Interaction Diagnostics}

Participants were instructed to consider primarily their experience of
speaking with and listening to \textsc{VoCa}:
\begin{enumerate}
  \setcounter{enumi}{16}
  \item \textsc{VoCa} generally understood what I intended to communicate.
  \item Across multiple turns, \textsc{VoCa} usually did not forget important
  information I had provided earlier.
  \item \textsc{VoCa}'s spoken responses were usually relevant to my question
  or request at the time.
  \item The waiting time before \textsc{VoCa} began responding was acceptable.
  \item I could tell when I could speak, wait, or interrupt.
  \item When \textsc{VoCa} misunderstood me, I could redirect the interaction
  toward my intended direction without much difficulty.
\end{enumerate}

\section{Semi-Structured Interview Questions}
\label{app:interview-guide}

\begin{enumerate}
  \item When do you typically use the voice agents you mentioned in
  the pre-study questionnaire? Please describe one representative experience.

  \item During the five days, in what situations did you use \textsc{VoCa},
  what were you trying to accomplish, and how was the overall experience?
  Which features or forms of interaction did you like most, which were most
  helpful, and why?

  \item Compared with the voice agents you had used before, what
  differences, if any, did \textsc{VoCa}'s combined Speech and Canvas responses
  make to your use? How did they affect what you were trying to accomplish?

  \item Please describe an occasion when \textsc{VoCa} did not meet your
  expectations. What were you trying to accomplish, how did it respond through
  Speech and the Canvas, and what did not meet your expectations?

  \item What capabilities would you like \textsc{VoCa} to add, or what existing
  features or interactions would you like it to improve? What problems would
  these changes address?
\end{enumerate}

\onecolumn
\section{Board Actions and Speech--Board Coordination}
\label{app:formative-coding}

\begin{table*}[h]
  \caption{Board-action types observed in Part~1. Roles describe uses in the
  discussion, rather than measured effects or exclusive action--need mappings.}
  \label{tab:formative-board-actions}
  \small
  \setlength{\tabcolsep}{5pt}
  \renewcommand{\arraystretch}{1.16}
  \begin{tabular}{@{}>{\raggedright\arraybackslash}p{0.045\textwidth}
    >{\raggedright\arraybackslash}p{0.19\textwidth}
    >{\raggedright\arraybackslash}p{\dimexpr0.465\textwidth-30pt\relax}
    >{\raggedright\arraybackslash}p{0.30\textwidth}@{}}
    \toprule
    Code & Board action & Description & Role in the discussion \\
    \midrule
    B1 & Recording information & Recording selected terms, quantities, or constraints as standalone content. & Keeping information available for discussion and later reference. \\
    B2 & Establishing a base structure & Drawing a basic figure, scene outline, grid, or structure around which discussion can develop. & Establishing a reference for objects, concepts, or spatial relationships. \\
    B3 & Adding detail & Adding labels, attributes, explanations, or enlarged detail views to existing content. & Elaborating structures, rules, or local details. \\
    B4 & Relating and grouping & Connecting or arranging elements to express relationships, sequence, membership, or grouping. & Making relationships and dependencies explicit. \\
    B5 & Emphasizing and marking status & Highlighting individual or grouped content, or marking priority, certainty, completion, or another status. & Directing shared attention and marking the status of ideas or decisions. \\
    B6 & Revising content & Correcting, updating, or removing information while largely retaining its overall organization. & Addressing errors, changed information, or revised constraints. \\
    B7 & Reorganizing and consolidating & Changing layout, hierarchy, or representation, or gathering dispersed content into an interim summary. & Reconsidering how ideas fit together and consolidating emerging outcomes. \\
    B8 & Pointing and demonstrating & Pointing, tracing along a diagram, or using bodily actions to explain board content. & Connecting speech to visible referents and demonstrating spatial or dynamic information. \\
    B9 & Simulating and calculating & Using the board to test assumptions, simulate a process, or calculate results, including non-numerical simulations. & Examining possible outcomes and the feasibility of proposals. \\
    \bottomrule
  \end{tabular}
\end{table*}

\begin{table*}[h]
  \caption{Speech--board coordination patterns in Part~1, ordered by decreasing session coverage.}
  \label{tab:formative-coordination}
  \small
  \setlength{\tabcolsep}{5pt}
  \renewcommand{\arraystretch}{1.16}
  \begin{tabular}{@{}>{\raggedright\arraybackslash}p{0.045\textwidth}
    >{\raggedright\arraybackslash}p{0.28\textwidth}
    >{\raggedright\arraybackslash}p{\dimexpr0.555\textwidth-30pt\relax}
    >{\raggedleft\arraybackslash}p{0.12\textwidth}@{}}
    \toprule
    Code & Coordination pattern & Speech--board relationship & Session coverage \\
    \midrule
    C1 & Synchronous speech and boardwork & Writing or drawing content concurrently with speech about that content. & 18/18 \\
    C2 & Visual updates prompted by questions or feedback & Adding, modifying, or reorganizing board content in response to a partner's spoken question, challenge, or feedback. & 17/18 \\
    C3 & Spoken reference to existing board content & Linking an utterance to a visible referent through deictic expressions coordinated with pointing. & 17/18 \\
    C4 & Returning to earlier board content in conversation & Resuming discussion of earlier board content while reviewing it to elaborate, check, or reassess prior information. & 12/18 \\
    C5 & Visual reorganization accompanying a discussion-stage transition & Reorganizing board content as talk moves into another stage, such as shifting from listing constraints to arranging a plan. & 10/18 \\
    C6 & Visual framework first & Drawing a basic framework before a spoken explanation unfolds around it. & 6/18 \\
    C7 & Spoken proposals for board content and organization & The non-operating partner verbally proposing how content should be drawn, classified, or arranged on the board. & 2/18 \\
    C8 & Selective or delayed board use & Continuing through speech without drawing, then beginning to draw when the discussion reaches structurally more complex content. & 1/18 \\
    C9 & Board-supported retelling and confirmation & Using board content to guide a spoken retelling, with the partner checking or confirming the account. & 1/18 \\
    \bottomrule
  \end{tabular}
\end{table*}

\end{document}